\documentclass[10pt,twocolumn,twoside]{IEEEtran}
\IfFileExists{algorithmic.sty}{\usepackage{algorithmic}}{}
\usepackage{graphicx}
\usepackage{tikz}
\usetikzlibrary{arrows.meta}
\usepackage{textcomp}
\def\BibTeX{{\rm B\kern-.05em{\sc i\kern-.025em b}\kern-.08em
    T\kern-.1667em\lower.7ex\hbox{E}\kern-.125emX}}
\usepackage{amsmath,amssymb,amsthm}
 \DeclareMathOperator{\rank}{rank}

\usepackage{color}
\IfFileExists{algorithm.sty}{\usepackage{algorithm}}{}

\usepackage{subeqnarray}
\AtBeginDocument{\setlength{\abovedisplayskip}{5pt plus 2pt minus 2pt}\setlength{\belowdisplayskip}{5pt plus 2pt minus 2pt}\setlength{\abovedisplayshortskip}{3pt plus 2pt minus 1pt}\setlength{\belowdisplayshortskip}{4pt plus 2pt minus 1pt}}

\IfFileExists{mathdots.sty}{\usepackage{mathdots}}{}
\newcommand{\safeincludegraphics}[2][]{%
  \IfFileExists{#2}{\includegraphics[#1]{#2}}{%
    \parbox{.8\columnwidth}{\centering Figure file unavailable:\\
      \texttt{\detokenize{#2}}}}%
}

\newtheorem{corollary}{Corollary}
\newtheorem{example}{Example}
\newtheorem{thm}{Theorem}

\newtheorem{corol}{Corollary}
\newtheorem{lemma}{Lemma}
\newtheorem{definition}{Definition}

\newtheorem{remark}{Remark}
\newcommand{\R}{\mathbb{R}}

\RequirePackage{color}\definecolor{RED}{rgb}{1,0,0}\definecolor{BLUE}{rgb}{0,0,1} 
\DeclareOldFontCommand{\sf}{\normalfont\sffamily}{\mathsf} 
\providecommand{\DIFdel}[1]{{}} 
\providecommand{\DIFaddbegin}{} 
\providecommand{\DIFaddend}{} 
\providecommand{\DIFdelbegin}{} 
\providecommand{\DIFdelend}{} 
\providecommand{\DIFaddbeginFL}{} 
\providecommand{\DIFaddendFL}{} 
\providecommand{\DIFdelbeginFL}{} 
\providecommand{\DIFdelendFL}{} 
\newcommand{\DIFscaledelfig}{0.5}
\RequirePackage{settobox} 
\RequirePackage{letltxmacro} 
\newsavebox{\DIFdelgraphicsbox} 
\newlength{\DIFdelgraphicswidth} 
\newlength{\DIFdelgraphicsheight} 
\LetLtxMacro{\DIFOincludegraphics}{\includegraphics} 
\newcommand{\DIFaddincludegraphics}[2][]{{\color{red}\fbox{\DIFOincludegraphics[#1]{#2}}}} 
\newcommand{\DIFdelincludegraphics}[2][]{
\sbox{\DIFdelgraphicsbox}{\DIFOincludegraphics[#1]{#2}}
\settoboxwidth{\DIFdelgraphicswidth}{\DIFdelgraphicsbox} 
\settoboxtotalheight{\DIFdelgraphicsheight}{\DIFdelgraphicsbox} 
\scalebox{\DIFscaledelfig}{
\parbox[b]{\DIFdelgraphicswidth}{\usebox{\DIFdelgraphicsbox}\\[-\baselineskip] \rule{\DIFdelgraphicswidth}{0em}}\llap{\resizebox{\DIFdelgraphicswidth}{\DIFdelgraphicsheight}{
\setlength{\unitlength}{\DIFdelgraphicswidth}
\begin{picture}(1,1)
\thicklines\linethickness{2pt} 
{\color[rgb]{1,0,0}\put(0,0){\framebox(1,1){}}}
{\color[rgb]{1,0,0}\put(0,0){\line( 1,1){1}}}
{\color[rgb]{1,0,0}\put(0,1){\line(1,-1){1}}}
\end{picture}
}\hspace*{3pt}}} 
} 
\LetLtxMacro{\DIFOaddbegin}{\DIFaddbegin} 
\LetLtxMacro{\DIFOaddend}{\DIFaddend} 
\LetLtxMacro{\DIFOdelbegin}{\DIFdelbegin} 
\LetLtxMacro{\DIFOdelend}{\DIFdelend} 
\DeclareRobustCommand{\DIFaddbegin}{\DIFOaddbegin \let\includegraphics\DIFaddincludegraphics} 
\DeclareRobustCommand{\DIFaddend}{\DIFOaddend \let\includegraphics\DIFOincludegraphics} 
\DeclareRobustCommand{\DIFdelbegin}{\DIFOdelbegin \let\includegraphics\DIFdelincludegraphics} 
\DeclareRobustCommand{\DIFdelend}{\DIFOaddend \let\includegraphics\DIFOincludegraphics} 
\LetLtxMacro{\DIFOaddbeginFL}{\DIFaddbeginFL} 
\LetLtxMacro{\DIFOaddendFL}{\DIFaddendFL} 
\LetLtxMacro{\DIFOdelbeginFL}{\DIFdelbeginFL} 
\LetLtxMacro{\DIFOdelendFL}{\DIFdelendFL} 
\DeclareRobustCommand{\DIFaddbeginFL}{\DIFOaddbeginFL \let\includegraphics\DIFaddincludegraphics} 
\DeclareRobustCommand{\DIFaddendFL}{\DIFOaddendFL \let\includegraphics\DIFOincludegraphics} 
\DeclareRobustCommand{\DIFdelbeginFL}{\DIFOdelbeginFL \let\includegraphics\DIFdelincludegraphics} 
\DeclareRobustCommand{\DIFdelendFL}{\DIFOaddendFL \let\includegraphics\DIFOincludegraphics} 
\RequirePackage{listings} 
\RequirePackage{color} 
\lstdefinelanguage{DIFcode}{ 
  moredelim=[il][\color{red}\scriptsize]{\%DIF\ <\ }, 
  moredelim=[il][\color{blue}\sffamily]{\%DIF\ >\ } 
} 
\lstdefinestyle{DIFverbatimstyle}{ 
	language=DIFcode, 
	basicstyle=\ttfamily, 
	columns=fullflexible, 
	keepspaces=true 
} 
\lstnewenvironment{DIFverbatim}{\lstset{style=DIFverbatimstyle}}{} 
\lstnewenvironment{DIFverbatim*}{\lstset{style=DIFverbatimstyle,showspaces=true}}{} 

\begin{document}

\parindent=1.25 em

\title{Strong Functional Observability and Strong Functional Detectability: Applications to Functional Observer Design for Systems with Unknown Inputs}
\smallskip
\author{M. Darouach 
\thanks{M. Darouach is with the Universit\'e de Lorraine, Centre de Recherche en Automatique de Nancy (CRAN UMR-7039, CNRS), IUT de Longwy, 186 rue de Lorraine 54400, Cosnes et Romain, France.
 (email: mohamed.darouach@univ-lorraine.fr)}
}
\maketitle
\begin{abstract}
This paper investigates strong functional observability, strong functional detectability, and strong$^*$ functional detectability for continuous-time linear systems with unknown inputs in both the state and measurement equations. Extending Kratz's finite algebraic approach \cite{26}, zero-output trajectories are characterized through the projected kernel of a finite matrix, without differentiating the unknown inputs. This characterization yields a necessary and sufficient rank condition for strong functional observability. For strong functional detectability, a frequency-dependent rank criterion involving only the original system matrices is derived, along with an equivalent finite test based on reduced zero-output dynamics. Strong$^*$ functional detectability is then characterized by an additional algebraic rank condition, and its relationship with the existence of stable functional observers is examined. A constructive observer design procedure based on constrained Sylvester equations is also developed. Particular cases and a numerical process example illustrate the theoretical results.
\end{abstract}

\section{Introduction}
Observability is a fundamental structural property of dynamical systems and plays a central role in estimation, control, monitoring, and fault diagnosis. Since the pioneering contributions of Kalman \cite{1} and Luenberger \cite{2,3}, observability theory has become a cornerstone of modern systems and control theory. Classical observability concerns the reconstruction of the complete state vector from available output measurements and provides the theoretical foundation for state observer design. In many practical applications, however, only a prescribed linear function $z(t)=Lx(t)$ needs to be estimated. Such a function may represent a performance variable, a physical quantity of interest, a fault indicator, or a combination of state variables used for monitoring and control. Direct estimation of these quantities can reduce the observer order and computational complexity, particularly in large-scale and interconnected systems, without requiring complete state observability \cite{4}-\cite{21}, \cite{29}, \cite{33}-\cite{34} .

These considerations have motivated the development of \emph{functional observability} and functional observer design, notably through the contributions of Fernando and coauthors \cite{7}--\cite{10}; see also \cite{15}-\cite{17}. Functional observability concerns the reconstruction of a prescribed function of the state rather than the entire state vector. Various algebraic and geometric characterizations, together with constructive observer design methods, have been developed in \cite{16}--\cite{19}. A closely related concept is \emph{functional detectability}, which ensures the asymptotic reconstruction of the prescribed linear function of the state \cite{16}, \cite{28}.

In practice, many engineering systems are affected by unknown disturbances, faults, or unmeasured inputs, which further complicate state estimation. In particular, unknown inputs may generate nonzero state trajectories while leaving the measured outputs identically zero. This difficulty has motivated the study of \emph{strong observability} and \emph{strong detectability} \cite{30}-\cite{32}. Hautus \cite{28} investigated strong detectability and its relationship with observer design, while Kratz \cite{26}, \cite{27} developed a finite algebraic construction for strong observability without requiring differentiability of the unknown inputs. These developments provide an important theoretical foundation for unknown-input observer (UIO) design \cite{35}, \cite{38}-\cite{41}, whose objective is to estimate the system state despite the presence of unknown disturbances. When only a prescribed function of the state is required, this leads naturally to \emph{functional unknown-input observers}. Their design has been investigated, in particular, by Darouach \cite{22}, Fernando and coauthors \cite{23}, and  \cite{25}.

Despite these developments, the structural relationships between functional observability, detectability, and unknown-input estimation remain less fully understood. A central question is to determine under which conditions a prescribed linear function of the state can be uniquely recovered or asymptotically determined in the presence of unknown inputs affecting both the state and measurement equations. Equally important is establishing how these conditions relate to the existence and design of stable functional observers.

Motivated by Kratz's approach \cite{26,27}, this paper develops finite algebraic characterizations of \emph{strong functional observability} (SFO), \emph{strong functional detectability} (SFD), and \emph{strong$^*$ functional detectability} (S$^*$FD) for continuous-time linear systems with unknown inputs. The proposed approach characterizes the initial states of admissible zero-output trajectories through the projected kernel of a  block matrix, without differentiating the unknown inputs, and yields a necessary and sufficient rank condition for SFO. For SFD, generalized coefficient chains lead to a frequency-dependent rank condition involving only the original system matrices, together with an equivalent finite computational test based on reduced zero-output dynamics. S$^*$FD is subsequently characterized by combining SFD with an additional algebraic rank condition. The relationships between these properties and stable functional observers are then examined, and a constructive observer design procedure based on constrained Sylvester equations is developed. The proposed approach extends the results of \cite{16}  to the analysis of strong functional observability and detectability, as well as to the design of functional observers for systems with unknown inputs.

Throughout this paper, $\mathbb R$ and $\mathbb C$ denote
the real and complex fields, respectively, and
${}^{\top}$ denotes transposition. The symbols $I_n$
and $0$ denote identity and zero matrices of appropriate
dimensions. For a matrix $M$, $\operatorname{rank}(M)$,
$\ker M$, $\operatorname{Im}M$, and $\operatorname{Row}(M)$
denote its rank, null space, column space, and row space,
respectively. 
The notation $\operatorname{col}(M_1,\ldots,M_h)$
denotes vertical concatenation and $e_i$ the $i$th canonical basis vector.
A generalized inverse $M^{-}$ satisfies $MM^{-}M=M$.
For $G\in\mathbb R^{p\times d}$, let
$G_\perp\in\mathbb R^{(p-\operatorname{rank}G)\times p}$
be a full-row-rank matrix satisfying
$G_\perp G=0$, so that
$\ker G_\perp=\operatorname{Im}G$.
For a square matrix $A$, $\sigma(A)$ denotes its
spectrum, $\operatorname{Re}(\lambda)$ the real part
of $\lambda$, and
$\mathbb C_-=\{\lambda\in\mathbb C:
\operatorname{Re}(\lambda)<0\}$.
A matrix is Hurwitz if its spectrum lies in
$\mathbb C_-$, while eigenvalues with nonnegative
real parts are called nonstable.
For $\lambda\in\sigma(A)$, $m_A(\lambda)$ denotes
its algebraic multiplicity, $A_\lambda=A-\lambda I_n$,
and $E_A(\lambda)=\ker A_\lambda^{m_A(\lambda)}$
the corresponding generalized eigenspace.
Spectral spaces and rank conditions involving
complex $\lambda$ are understood over $\mathbb C$.
\section{Preliminaries}
We first introduce the system model and the finite Kratz construction used throughout the paper. \\
Consider the continuous-time linear system
\begin{subequations}\label{sys-kratz}
\begin{align}
\dot{x}(t) &= Ax(t)+Bu(t)+Fd(t),
\label{sys-kratza}\\
y(t) &= Cx(t)+Gd(t),
\label{sys-kratzb}\\
z(t) &= Lx(t),
\label{sys-kratzc}
\end{align}
\end{subequations}
with initial state $x(0)\in\mathbb R^n$.
The system matrices are real and have the following
dimensions: $A\in\mathbb{R}^{n\times n}$,
$B\in\mathbb{R}^{n\times m}$,
$F\in\mathbb{R}^{n\times d}$,
$C\in\mathbb{R}^{p\times n}$,
$G\in\mathbb{R}^{p\times d}$, and
$L\in\mathbb{R}^{r\times n}$.
We assume that the unknown input
$d(t)\in L^1_{\mathrm{loc}}([0,\infty);\mathbb R^d)$ and that
the corresponding state trajectory $x(t)$ is continuous.
Let $M\in\mathbb R^{s\times n}$ and
$\Gamma\in\mathbb R^{s\times d}$.
For each integer $h\geq1$, define
$
\mathcal O_{(M,A,h)}
=
\begin{bmatrix}
M\\
MA\\
\vdots\\
MA^{h-1}
\end{bmatrix}
$
and the lower block-triangular matrix
\[
\mathcal T_{(M,F,\Gamma,h)}
=
\begin{bmatrix}
\Gamma&0&\cdots&0\\
MF&\Gamma&\ddots&\vdots\\
\vdots&\ddots&\ddots&0\\
MA^{h-2}F&\cdots&MF&\Gamma
\end{bmatrix}.
\]
More precisely, its $(i,j)$ block,
$0\leq i,j<h$, is
\[
\bigl(\mathcal T_{(M,F,\Gamma,h)}\bigr)_{ij}
=
\begin{cases}
MA^{i-j-1}F,&j<i,\\
\Gamma,&j=i,\\
0,&j>i.
\end{cases}
\]
Define
\begin{equation}\label{eq:Kratz-general}
\mathcal S_{(M,F,\Gamma,h)}
=
\begin{bmatrix}
\mathcal O_{(M,A,h)}&
\mathcal T_{(M,F,\Gamma,h)}
\end{bmatrix}.
\end{equation}
The dimensions are
\[
\begin{aligned}
\mathcal O_{(M,A,h)}
&\in\mathbb R^{hs\times n},\\
\mathcal T_{(M,F,\Gamma,h)}
&\in\mathbb R^{hs\times hd},\\
\mathcal S_{(M,F,\Gamma,h)}
&\in\mathbb R^{hs\times(n+hd)}.
\end{aligned}
\]
In particular,
$
\mathcal S_{(C,F,G,1)}
=\begin{bmatrix}C&G\end{bmatrix},
$
and
$
\mathcal S_{(C,F,G,2)}
=\begin{bmatrix}
C&G&0\\
CA&CF&G
\end{bmatrix}.
$
The notions of SFO, SFD, and S$^*$FD are considered
for system \eqref{sys-kratz} with $u(t)=0$.
The corresponding zero-output dynamics are governed by
\begin{equation}\label{eq:zero-output-ct}
\dot{x}(t)=Ax(t)+Fd(t),\qquad
Cx(t)+Gd(t)=0.
\end{equation}
Our objective is to determine, through finite matrix
computations, whether zero output implies that $Lx(t)$
vanishes identically or converges to zero as $t\to\infty$.
Following Kratz's approach \cite{26}, consider the
finite algebraic relations associated with
\eqref{eq:zero-output-ct}:
\begin{equation}\label{eq:Kratz-finite-rec}
\begin{aligned}
x_{j+1}&=Ax_j+Fd_j,\\
0&=Cx_j+Gd_j,
\end{aligned}
\qquad j=0,\ldots,h-1.
\end{equation}
Here, $x_j\in\mathbb R^n$ and
$d_j\in\mathbb R^d$ are algebraic
Kratz coefficients. The Kratz construction
is not a time discretization of the
continuous-time system, and the index $j$
does not represent a discrete-time instant.
In particular, $x_0$ denotes the first
algebraic coefficient and should not be
confused with the initial state $x(0)$
of an arbitrary trajectory. The equality
$x(0)=x_0$ holds only when a trajectory
with this initial state is explicitly
constructed. From \eqref{eq:Kratz-finite-rec},
we obtain
\begin{equation}\label{eq:Kratz-state-expansion}
 x_j=A^jx_0+\sum_{i=0}^{j-1}A^{j-1-i}Fd_i.
\end{equation}
Substituting into the zero-output constraint yields
\begin{equation}\label{eq:Kratz-output-expansion}
CA^jx_0+\sum_{i=0}^{j-1}CA^{j-1-i}Fd_i
+Gd_j=0.
\end{equation}
Introducing
$\chi_h=\operatorname{col}(x_0,d_0,\ldots,d_{h-1})$,
the equations \eqref{eq:Kratz-output-expansion}
can be written in the compact form
\begin{equation}\label{eq:Kratz-stack-simple}
\mathcal S_{(C,F,G,h)}\chi_h=0.
\end{equation}
The following lemma establishes the equivalence between equation~\eqref{eq:Kratz-stack-simple} and the algebraic chain~\eqref{eq:Kratz-finite-rec}.
\begin{lemma}
\label{lem:Kratz-chain-equivalence}
For a prescribed $x_0\in\mathbb R^n$ and $h\geq1$,
there exist $d_0,\ldots,d_{h-1}$ satisfying
\eqref{eq:Kratz-stack-simple} if and only if there exist
$x_1,\ldots,x_h$ and $d_0,\ldots,d_{h-1}$ satisfying
\eqref{eq:Kratz-finite-rec}. For each fixed
$(x_0,d_0,\ldots,d_{h-1})$, the algebraic coefficients
$x_1,\ldots,x_h$ are uniquely determined.
\end{lemma}
\begin{proof}
Suppose first that \eqref{eq:Kratz-finite-rec}
holds. Successive substitution of the state
equations gives \eqref{eq:Kratz-state-expansion}.
The zero-output equations then yield
\eqref{eq:Kratz-output-expansion}, which is
equivalent to \eqref{eq:Kratz-stack-simple}.

Conversely, we proceed by induction on $h$.
For $h=1$, the equation
$\mathcal S_{(C,F,G,1)}\chi_1=0$
reduces to
$
Cx_0+Gd_0=0.
$
Taking $x_1=Ax_0+Fd_0$ gives both
equations in \eqref{eq:Kratz-finite-rec}.
For $h=2$, the equation
$\mathcal S_{(C,F,G,2)}\chi_2=0$
gives
$
Cx_0+Gd_0=0,\qquad
CAx_0+CFd_0+Gd_1=0.
$
Introduce the algebraic coefficient
$x_1=Ax_0+Fd_0$.
The second equation then becomes
$
Cx_1+Gd_1=0.
$
Introducing $x_2=Ax_1+Fd_1$
completes the algebraic chain for $h=2$.
Now suppose that the equivalence holds
for some $h\geq2$, and consider
$
\mathcal S_{(C,F,G,h+1)}
\chi_{h+1}=0.
$
Its first $h$ block rows give
$
\mathcal S_{(C,F,G,h)}
\operatorname{col}(x_0,d_0,\ldots,d_{h-1})
=0.
$
By the induction hypothesis, there
exist uniquely determined coefficients
$x_1,\ldots,x_h$ satisfying
$
x_{j+1}=Ax_j+Fd_j,\qquad
Cx_j+Gd_j=0,\quad 0\leq j<h.
$
The last block row of
$\mathcal S_{(C,F,G,h+1)}\chi_{h+1}=0$
is
$
CA^hx_0+
\sum_{i=0}^{h-1}CA^{h-1-i}Fd_i
+Gd_h=0.
$
Using \eqref{eq:Kratz-state-expansion},
this becomes
$
Cx_h+Gd_h=0.
$
Introducing the algebraic coefficient
$x_{h+1}=Ax_h+Fd_h$ completes the chain.
Thus, the equivalence holds for $h+1$,
and the proof follows by induction.
\end{proof}
For $h\geq 1$, define
$
\Pi_{0,h}=
\begin{bmatrix}I_n&0_{n\times hd}\end{bmatrix},
\qquad
\mathcal X_h=
\Pi_{0,h}\ker\mathcal S_{(C,F,G,h)},
$
and set $\mathcal X_0=\mathbb R^n$.
For $h=n$,   set  $\Pi_0=\Pi_{0,n}$ and   let
 \begin{equation}\label{eq:Kratz-V}
\operatorname{Im}V
=\mathcal X_n
=\Pi_0\ker\mathcal S  _{(C,F,G,n)} ,
\end{equation}
where $V\in\mathbb R^{n\times w}$ has full column
rank and $w=\dim\mathcal X_n$.
Thus, a basis of $\mathcal X_n$ is obtained
from the null space of a single  matrix.
For each   $h\geq0$, choose a full-row-rank matrix
$Q_h\in\mathbb R^{(n-\dim\mathcal X_h)\times n}$
such that  $\mathcal X_h=\ker Q_h$.  In particular,
$\operatorname{Im}V=\mathcal X_n=\ker Q_n$.
The following lemma establishes the properties
of the subspaces $\mathcal X_h$ and their
relationship with admissible zero-output
trajectories.
\begin{lemma}
\label{lem:Kratz-SFD-completeness}
The following properties hold:
\begin{enumerate}
\item The spaces $\mathcal X_h$ satisfy
$
\mathcal X_0\supseteq\mathcal X_1
\supseteq\cdots\supseteq
\mathcal X_n=\mathcal X_{n+1}=\cdots.
$
\item Every admissible zero-output trajectory
on $[0,T]$ satisfies
$
x(t)\in\mathcal X_n,
\qquad t\in[0,T].
$
\item Every $v\in\mathcal X_n$ is the
initial state of an admissible zero-output
trajectory. Moreover, there exist matrices
$K\in\mathbb R^{d\times w}$ and
$A_h\in\mathbb R^{w\times w}$,
where $w=\operatorname{rank}V$, such that
\begin{equation}\label{eq:Kratz-SFD-invariance}
CV+GK=0,\qquad AV+FK=VA_h.
\end{equation}
\end{enumerate}
\end{lemma}
\begin{proof}
\emph{1)}
Recall that
$
\mathcal X_h
=\Pi_{0,h}\ker\mathcal S_{(C,F,G,h)}.
$
By the definition of the projected kernel and
Lemma~\ref{lem:Kratz-chain-equivalence},
a vector $v$ belongs to $\mathcal X_{h+1}$
if and only if there exist algebraic
coefficients $x_0,\ldots,x_{h+1}$
and $d_0,\ldots,d_h$ satisfying
\[
\begin{aligned}
x_0&=v,\\
x_{j+1}&=Ax_j+Fd_j,\\
Cx_j+Gd_j&=0,\qquad j=0,\ldots,h.
\end{aligned}
\]
The equations for $j=0$ give
$
Cv+Gd_0=0,\qquad x_1=Av+Fd_0.
$
The remaining equations, for
$j=1,\ldots,h$, form a chain of length $h$
with initial coefficient $x_1$; by
Lemma~\ref{lem:Kratz-chain-equivalence} and the
projected-kernel definition, they are equivalent to
$x_1\in\mathcal X_h$. Conversely, such a chain
can be preceded by the pair $(v,d_0)$ whenever
$Cv+Gd_0=0$ and $x_1=Av+Fd_0$. Therefore,
\[
\mathcal X_{h+1}
=\left\{v\in\mathbb R^n:
\begin{array}{l}
\exists\,\delta\in\mathbb R^d
\text{ such that}\\
Cv+G\delta=0,\\
Av+F\delta\in\mathcal X_h
\end{array}
\right\}.
\]
Any Kratz chain of length $h+1$
contains a chain of length $h$, which implies
 $\mathcal X_{h+1}\subseteq\mathcal X_h$.
Moreover, the recursion shows that
if $\mathcal X_{h+1}=\mathcal X_h$,
then $\mathcal X_{h+2}=\mathcal X_{h+1}$.
Since $\mathcal X_0=\mathbb R^n$,
there can be at most $n$ strict
inclusions. Therefore,
$
\mathcal X_n=\mathcal X_{n+1}
=\cdots.
$

\emph{2)}
Let $(x(t),d(t))$ be an admissible
zero-output trajectory on $[0,T]$.
We prove by induction that
$x(t)\in\mathcal X_h$ for every $h$.
The result is immediate for $h=0$. Suppose that $x(t)\in\mathcal X_h$ for all
$t\in[0,T]$, and choose $Q_h$ such that
$\ker Q_h=\mathcal X_h$.
Since $Q_hx(t)=0$, the integral state equation gives
$\int_s^t Q_h(Ax(\tau)+Fd(\tau))\,d\tau=0$
for every $0\leq s\leq t\leq T$.
Thus $Q_h(Ax+Fd)=0$ along the trajectory wherever the
state equation holds. Together with the zero-output relation,
the recursion in part 1 gives $x(t)\in\mathcal X_{h+1}$.
By continuity of $x$ and closedness of $\mathcal X_{h+1}$,
this inclusion holds for every $t\in[0,T]$. Consequently,
$
x(t)\in\mathcal X_n,\qquad t\in[0,T].
$

\emph{3)}
Since $\mathcal X_n=\mathcal X_{n+1}$,
the recursion shows that for every
$v\in\mathcal X_n$, there exists
$\delta\in\mathbb R^d$ such that
\[
Cv+G\delta=0,\qquad
Av+F\delta\in\mathcal X_n.
\]
Applying this property to each
column $v_i$ of $V$, we obtain
vectors $k_i\in\mathbb R^d$ and
$a_i\in\mathbb R^w$ satisfying
\[
Cv_i+Gk_i=0,\qquad
Av_i+Fk_i=Va_i.
\]
Collecting these vectors into
$K$ and $A_h$ gives
\eqref{eq:Kratz-SFD-invariance}.
Now, let $v\in\mathcal X_n$
and choose $\eta(0)$ such that
$v=V\eta(0)$. Define
\begin{equation}\label{eq:Kratz-SFD-construction}
\begin{aligned}
x(t)&=Ve^{A_ht}\eta(0),\\
d(t)&=Ke^{A_ht}\eta(0).
\end{aligned}
\end{equation}
Using \eqref{eq:Kratz-SFD-invariance},
we obtain
$
\dot x(t)=VA_he^{A_ht}\eta(0)=(AV+FK)e^{A_ht}\eta(0)=Ax(t)+Fd(t),
$
and
$
Cx(t)+Gd(t)
=(CV+GK)e^{A_ht}\eta(0)=0.
$
Thus, $(x(t),d(t))$ is an admissible
zero-output trajectory satisfying
$x(0)=v$. For $w=0$, the zero
trajectory gives the same conclusion.
\end{proof}
Lemma~\ref{lem:Kratz-SFD-completeness} shows that
$\mathcal X_n$ characterizes exactly the
initial states of admissible zero-output trajectories
and contains their states at all times. This result
provides the basis for the subsequent characterization
of SFO and SFD.
The following example illustrates Lemma~\ref{lem:Kratz-SFD-completeness}.
\begin{example}
Consider the system described by the following matrices:
$
A=\begin{pmatrix}
-3&1&0\\
0&-3&1\\
1&0&-3
\end{pmatrix},$
\quad
$F=\begin{pmatrix}1\\1\\1\end{pmatrix},
$
$
C=\begin{pmatrix}
-1&0&0\\
1&-1&0
\end{pmatrix}
$
\quad \text{and} \quad
$G=\begin{pmatrix}1\\0\end{pmatrix}.$
For $h=1$, the equations
$\mathcal S_{(C,F,G,1)}\chi_1=0$
give $d_0=x_{0,1}$ and $x_{0,2}=x_{0,1}$.
Then
$
\mathcal X_1
=\Pi_{0,1}\ker\mathcal S_{(C,F,G,1)}
=\operatorname{span}\{(1,1,0)^\top,e_3\}.
$
For $h=2$, write $x_0=(a,a,b)^\top$
and $d_0=a$. The state recursion gives
$
x_1=Ax_0+Fd_0
=(-a,b-2a,2a-3b)^\top.
$
The second component of $Cx_1+Gd_1=0$
requires $x_{1,1}=x_{1,2}$, then $b=a$.
Therefore
$
\mathcal X_2
=\Pi_{0,2}\ker\mathcal S_{(C,F,G,2)}
=\operatorname{span}\{(1,1,1)^\top\}.
$
For $x_0=a(1,1,1)^\top$, the choice
$d_0=a$ gives $x_1=-a(1,1,1)^\top$;
choosing $d_1=-a$ satisfies the next
output equation. Thus
$\mathcal X_3=\mathcal X_2$.
With $\Pi_0=[\,I_3\ \ 0_{3\times3}\,]$,
we obtain
$
\Pi_0\ker\mathcal S_{(C,F,G,3)}
=\operatorname{Im}V,\qquad
V=(1,1,1)^\top.
$
By solving the following equations
$
CV+GK=0,\qquad AV+FK=VA_h.
$
we obtain $K=1$ and $A_h=-1$.
Consequently, for every $a\in\mathbb R$,
$
x(t)=ae^{-t}V,\qquad d(t)=ae^{-t}
$
is an admissible zero-output trajectory
with $x(0)=aV$. For $a\neq0$, all three
state components and the unknown input
are nonzero.
\end{example}
This example illustrates
Lemma~\ref{lem:Kratz-SFD-completeness}:
the subspaces $\mathcal X_h$, obtained
from the finite Kratz construction,
stabilize at $\operatorname{Im}V$.
Moreover, every initial state in \(\operatorname{Im}V\) generates an admissible zero-output trajectory, as established in Lemma~\ref{lem:Kratz-SFD-completeness}.

We close the preliminaries with a standard linear-algebraic criterion used repeatedly in the sequel (see \cite{24} and references therein).
Consider the following consistent algebraic equations for the unknown vectors $x$ and $z$.
\begin{subeqnarray}\label{eq11}
  &&\mathcal{A}x=b, 
        \slabel{eq11a}\\
  &&z=\mathcal{L}x,   \slabel{eq11b}
\end{subeqnarray}
where $x\in \R^n$, $z\in \R^{r}$ and $b\in \mathbb{R}^{m}$ with $\mathcal A\in\mathbb R^{m\times n}$ and $\mathcal L\in\mathbb R^{r\times n}$.
Let $\mathcal S=\{x\in\mathbb R^n:\mathcal A x=b\}\neq\emptyset$ be the solution set.
\begin{lemma}\label{lem2}  
The following assertions are equivalent:
\begin{enumerate}
\item The functional $z$ is uniquely determined from \eqref{eq11a}.
\item $\operatorname{Row}(\mathcal L)\subseteq\operatorname{Row}(\mathcal A)$.
\item $\rank\begin{pmatrix}\mathcal{L}\\\mathcal{A}\end{pmatrix}=\rank (\mathcal{A})$.
\item $\ker(\mathcal{A}) \subseteq \ker(\mathcal{L})$.
\end{enumerate}
\end{lemma}
\begin{proof}
Let \(x_*\) be one solution of the consistent equations
\(\mathcal A x=b\). Every solution is \(x=x_*+v\),
where \(v\in\ker\mathcal A\). Therefore
\(\mathcal Lx=\mathcal Lx_*+\mathcal Lv\) is
independent of the solution exactly when
\(\ker\mathcal A\subseteq\ker\mathcal L\).
The orthogonal-complement identity
\((\ker\mathcal A)^\perp=\operatorname{Row}(\mathcal A)\)
makes this equivalent to inclusion of row spaces. Adding
the rows of \(\mathcal L\) then does not increase the
rank, proving the equivalence with the stacked rank condition.
Conversely, equality of the ranks means that each added row
is a linear combination of the original rows, so
\(\mathcal L=\Omega\mathcal A\) for some \(\Omega\).
This also gives the unique functional value
\(\mathcal Lx=\Omega b\).
\end{proof}
\section{Strong functional observability, Strong functional detectability and Strong$^*$ functional detectability}
This section considers SFO, SFD, and S$^*$FD for system \eqref{sys-kratz} with $u(t)=0$, namely, 
\begin{subequations}\label{sys-kratz-md}
\begin{align}
\dot{x}(t) &= Ax(t)+Fd(t),
\label{sys-kratzmda}\\
y(t) &= Cx(t)+Gd(t),
\label{sys-kratzmdb}\\
z(t) &= Lx(t),
\label{sys-kratzmdc}
\end{align}
\end{subequations}
\subsection{Strong functional observability}
We first define the SFO.
\begin{definition}\label{def1}
System \eqref{sys-kratz-md} is strongly
functionally observable with respect to
$z(t)=Lx(t)$ if, for every admissible unknown
input $d(t)$, every initial state, and every
$T>0$,
\[
y(t)=0 \text{ on }[0,T]
\quad\Longrightarrow\quad
Lx(t)=0 \text{ for all }t\in[0,T].
\]
\end{definition}
Let $\mathcal L_0=
\begin{bmatrix}
L&0_{r\times nd}
\end{bmatrix}.$

The following theorem provides a necessary and sufficient
condition for the SFO of
system~\eqref{sys-kratz-md}.

\begin{thm}\label{thm3MD}
System~\eqref{sys-kratz-md} is strongly functionally
observable with respect to $z(t)=Lx(t)$ if and only if
\begin{equation}\label{eq:SFO-rank-simple}
\rank
\begin{bmatrix}
\mathcal S_{(C,F,G,n)}\\
\mathcal L_0
\end{bmatrix}
=\rank \mathcal S_{(C,F,G,n)}.
\end{equation}
\end{thm}

\begin{proof}
By Lemma~\ref{lem:Kratz-SFD-completeness},
every admissible zero-output trajectory
satisfies $x(t)\in\mathcal X_n$, and every
$v\in\mathcal X_n$ is realizable as
$x(0)=v$ along such a trajectory.
Therefore, SFO
is equivalent to
\[
\mathcal X_n\subseteq\ker L.
\]
Since
$\mathcal X_n=\Pi_0\ker\mathcal S_{(C,F,G,n)}$,
this inclusion is equivalent to
\[
\ker\mathcal S_{(C,F,G,n)}
\subseteq\ker(L\Pi_0).
\]
Using $\mathcal L_0=L\Pi_0$, the
kernel--rank equivalence of Lemma~3
gives
\[
\operatorname{rank}
\begin{bmatrix}
\mathcal S_{(C,F,G,n)}\\
\mathcal L_0
\end{bmatrix}
=
\operatorname{rank}\mathcal S_{(C,F,G,n)},
\]
which completes the proof.
\end{proof}
\color{black}

\subsection*{Particular cases of SFO}
Now, we can apply the result of Theorem \ref{thm3MD} to some particular cases:
\begin{corol}
For $L=I_n$ and $G=0$, let
$\mathcal S_n=\mathcal S_{(C,F,0,n)}$ 
Then the SFO condition of Theorem~\ref{thm3MD} reduces to
\begin{equation}\label{eq:Kratz-recovered}
\rank\begin{bmatrix}\mathcal S_n\\\Pi_0\end{bmatrix}
=\rank\mathcal S_n,
\end{equation}
which yields the strong-observability rank criterion established by Kratz~\cite26.
\end{corol}
\begin{corol}
If $F=0$ and $G=0$, the system is  functionally observable with respect to
$Lx(t)$ if and only if
\begin{equation}\label{eq:autonomous-FO-rank}
\rank\begin{pmatrix}\mathcal O(C,A,n)\\L\end{pmatrix}
=\rank\mathcal O(C,A,n),
\end{equation}
or, equivalently,
$\ker\mathcal O(C,A,n)\subseteq\ker L$.
\end{corol}
This recovers the functional observability criterion in
\cite{8}, \cite{9} and \cite{17}.
\begin{corol}
\label{cor:autonomous-Kalman}
If $F=0$, $G=0$ and in addition, $L=I_n$, SFO is equivalent to
\begin{equation}\label{eq:Kalman-explicit}
\rank\mathcal O(C,A,n)=n,
\end{equation}
the classical Kalman observability condition.
\end{corol}
\subsection{Strong functional detectability}
SFO requires the
functional to vanish whenever the measured
output is zero. SFD
requires only that the functional converge to
zero along every zero-output trajectory.
\begin{definition}\label{def:SFD}
System \eqref{sys-kratz-md} is strongly
functionally detectable with respect to
$z(t)=Lx(t)$ if every admissible trajectory
satisfying $y(t)=0$ on
$[0,\infty)$ also satisfies
$
\lim_{t\to\infty}Lx(t)=0.
$
\end{definition}
Before establishing the necessary and sufficient
conditions for SFD of
system~\eqref{sys-kratz-md}, we introduce some
preliminary algebraic constructions.

Recall that $A_\lambda=A-\lambda I_n$.
For $\lambda\in\mathbb C$ and $k\geq1$,
define
\begin{equation}\label{eq:SFD-TF}
\begin{aligned}
\mathcal T_k(\lambda)
&=
\begin{bmatrix}
\mathcal O_k(\lambda)&\mathcal H_k(\lambda)\\
A_\lambda^k&\mathcal P_k(\lambda)
\end{bmatrix},\\
\mathcal F_k(\lambda)
&=
\begin{bmatrix}
\mathcal O_k^L(\lambda)&
\mathcal H_k^L(\lambda)
\end{bmatrix},
\end{aligned}
\end{equation}
where
\begin{equation}\label{eq:SFD-OP}
\begin{aligned}
\mathcal O_k(\lambda)
&=
\begin{bmatrix}
C\\
CA_\lambda\\
\vdots\\
CA_\lambda^{k-1}
\end{bmatrix},\\[1mm]
\mathcal O_k^L(\lambda)
&=
\begin{bmatrix}
L\\
LA_\lambda\\
\vdots\\
LA_\lambda^{k-1}
\end{bmatrix},\\[1mm]
\mathcal P_k(\lambda)
&=
\begin{bmatrix}
A_\lambda^{k-1}F&
A_\lambda^{k-2}F&
\cdots&
F
\end{bmatrix}.
\end{aligned}
\end{equation}
Thus, $\mathcal O_k(\lambda)$ and
$\mathcal O_k^L(\lambda)$ are block-column
matrices, whereas $\mathcal P_k(\lambda)$
is a block-row matrix.

The matrices $\mathcal H_k(\lambda)$ and
$\mathcal H_k^L(\lambda)$ are lower
block-triangular matrices. Their $(i,j)$
blocks, indexed from $0$ to $k-1$, are
defined by
\begin{equation}\label{eq:SFD-H-blocks}
\begin{aligned}
(\mathcal H_k(\lambda))_{ij}
&=
\begin{cases}
CA_\lambda^{i-j-1}F,&j<i,\\
G,&j=i,\\
0,&j>i,
\end{cases}\\[1mm]
(\mathcal H_k^L(\lambda))_{ij}
&=
\begin{cases}
LA_\lambda^{i-j-1}F,&j<i,\\
0,&j\geq i.
\end{cases}
\end{aligned}
\end{equation}
In particular,
\[
\mathcal H_k(\lambda)=
\begin{bmatrix}
G&0&\cdots&0\\
CF&G&\ddots&\vdots\\
CA_\lambda F&CF&\ddots&0\\
\vdots&\vdots&\ddots&G
\end{bmatrix},
\]
and
\[
\mathcal H_k^L(\lambda)=
\begin{bmatrix}
0&0&\cdots&0\\
LF&0&\ddots&\vdots\\
LA_\lambda F&LF&\ddots&0\\
\vdots&\vdots&\ddots&0
\end{bmatrix}.
\]
The dimensions of these matrices are
\[
\begin{aligned}
\mathcal O_k&\in\mathbb C^{kp\times n},
&\mathcal O_k^L&\in\mathbb C^{kr\times n},\\
\mathcal H_k&\in\mathbb C^{kp\times kd},
&\mathcal H_k^L&\in\mathbb C^{kr\times kd},\\
\mathcal P_k&\in\mathbb C^{n\times kd}.
\end{aligned}
\]
Consequently,
\[
\begin{aligned}
\mathcal T_k(\lambda)
&\in\mathbb C^{(kp+n)\times(n+kd)},\\
\mathcal F_k(\lambda)
&\in\mathbb C^{kr\times(n+kd)}.
\end{aligned}
\]

The following lemma establishes an exact
reduced representation of the zero-output
dynamics, which will be used to characterize
SFD.

\begin{lemma}
\label{prop:exact-zero-output-reduction}
Let $V$, $K$, and $A_h$ be defined as in
Lemma~\ref{lem:Kratz-SFD-completeness}.
Consider the vector subspace
\[
\mathcal U_h=
\{\delta\in\mathbb R^d:
G\delta=0,\ F\delta\in\operatorname{Im}V\},
\]
and let the columns of
$D_h\in\mathbb R^{d\times\ell_h}$ form a basis
of $\mathcal U_h$, where
$\ell_h=\dim\mathcal U_h$.
Since $\operatorname{Im}(FD_h)
\subseteq\operatorname{Im}V$, there exists
$B_h\in\mathbb R^{w\times\ell_h}$ such that
\[
FD_h=VB_h.
\]
Set $C_h=LV$.

Then every admissible zero-output trajectory
admits the representation
\[
x(t)=V\eta(t),\qquad
d(t)=K\eta(t)+D_h\nu(t),
\]
where $\eta$ is continuous,
$\nu\in L^1_{\mathrm{loc}}([0,\infty);\mathbb R^{\ell_h})$, and
\begin{equation}\label{eq:SFD-reduced}
\dot\eta(t)=A_h\eta(t)+B_h\nu(t),
\qquad z(t)=C_h\eta(t).
\end{equation}
Conversely, every solution of
\eqref{eq:SFD-reduced}, together with the
above representation of $x$ and $d$,
defines an admissible zero-output trajectory
of the original system.
\end{lemma}
\begin{proof}
By Lemma~\ref{lem:Kratz-SFD-completeness},
every zero-output trajectory satisfies
$x(t)\in\operatorname{Im}V$ and can therefore
be written as $x(t)=V\eta(t)$, with $\eta$ continuous.
Using the system equations and
$CV+GK=0$, $AV+FK=VA_h$, we obtain
\[
\begin{aligned}
G(d-K\eta)&=0,\\
F(d-K\eta)&=V(\dot\eta-A_h\eta).
\end{aligned}
\]
Then $d-K\eta\in\mathcal U_h$, so that
$d=K\eta+D_h\nu$ for some
$\nu\in L^1_{\mathrm{loc}}$.
Substitution yields
$
V\dot\eta=V(A_h\eta+B_h\nu).
$
Since $V$ has full column rank,
$
\dot\eta=A_h\eta+B_h\nu,
\qquad z=LV\eta=C_h\eta.
$
Conversely, let $\eta,\nu$ satisfy
\eqref{eq:SFD-reduced} and define
$x=V\eta$, $d=K\eta+D_h\nu$.
The identities
$
VA_h=AV+FK,\quad VB_h=FD_h,
\quad CV+GK=0,\quad \text{and} \quad GD_h=0
$
give
$
\dot x=Ax+Fd,\qquad Cx+Gd=0.
$
Thus, the reduced system provides an exact
representation of all admissible
zero-output trajectories.
\end{proof}
%
\begin{remark}
The reduced system matrices can be computed by
first determining $V$ from \eqref{eq:Kratz-V},
then solving
\begin{equation}\label{eq:SFD-reduced-matrices}
\begin{aligned}
CV+GK&=0, & AV+FK&=VA_h,\\
GD_h&=0, & FD_h&=VB_h,\\
C_h&=LV,
\end{aligned}
\end{equation}
where the columns of $D_h$ form a basis of
$\mathcal U_h$. The procedure is summarized
in Algorithm~1.

If $\operatorname{rank}V=0$, SFD holds
automatically. Otherwise, the reduced system
\eqref{eq:SFD-reduced} can be used to test SFD.
\end{remark}

The following theorem provides necessary and sufficient conditions for SFD.

\begin{thm}\label{thm:SFD-direct}
The following statements are equivalent:
\begin{enumerate}
\item System~\eqref{sys-kratz-md} is strongly
functionally detectable with respect to
$z(t)=Lx(t)$.

\item For every $\lambda\in\mathbb C$ with
$\operatorname{Re}\lambda\geq0$,
\begin{equation}\label{eq:SFD-chain}
\rank
\begin{bmatrix}
\mathcal T_n(\lambda)\\
\mathcal F_n(\lambda)
\end{bmatrix}
=\rank\mathcal T_n(\lambda).
\end{equation}

\item The reduced system \eqref{eq:SFD-reduced}
satisfies
\begin{align}
C_h\mathcal R_h&=0,
\label{eq:SFD-reduced-condition1}\\
\rank
\begin{bmatrix}
(A_h-\lambda I_w)^{m_h(\lambda)}\\
C_h
\end{bmatrix}
&=\rank(A_h-\lambda I_w)^{m_h(\lambda)},
\label{eq:SFD-reduced-condition2}
\end{align}
for every $\lambda\in\sigma(A_h)$ with
$\operatorname{Re}\lambda\geq0$, where
\[
\mathcal R_h=
[\,B_h\ A_hB_h\ \cdots\ A_h^{w-1}B_h\,],
\]
and $m_h(\lambda)$ denotes the algebraic
multiplicity of $\lambda$ in $A_h$.
\end{enumerate}
\end{thm}

\begin{proof}

\emph{(1) $\Rightarrow$ (2).}

Assume SFD and let $\lambda\in\mathbb C$ satisfy
$\operatorname{Re}\lambda\geq0$.
Consider any
$\chi_n=\operatorname{col}(x_0,d_0,\ldots,d_{n-1})
\in\ker\mathcal T_n(\lambda)$.

Starting from $x_0$, define recursively
$
x_{j+1}=A_\lambda x_j+Fd_j,
\qquad 0\leq j<n.
$
Successive substitution gives
$
x_j=A_\lambda^jx_0+
\sum_{i=0}^{j-1}A_\lambda^{j-1-i}Fd_i,
\qquad 1\leq j\leq n.
$
By the block structure of $\mathcal T_n(\lambda)$,
its first $n$ block rows are precisely
\[
Cx_j+Gd_j=0,\qquad 0\leq j<n,
\]
while its last block row gives
$
x_n=A_\lambda^nx_0+
\sum_{i=0}^{n-1}A_\lambda^{n-1-i}Fd_i=0.
$
We now show that these algebraic relations
yield a zero-output trajectory of the
original system. To this end, consider
\begin{equation}\label{eq:chain-trajectory}
\begin{aligned}
x(t)&=e^{\lambda t}
\sum_{j=0}^{n-1}\frac{t^j}{j!}x_j,\\
d(t)&=e^{\lambda t}
\sum_{j=0}^{n-1}\frac{t^j}{j!}d_j.
\end{aligned}
\end{equation}
Differentiating $x(t)$ and using $x_n=0$
and $A_\lambda=A-\lambda I_n$, we obtain
\[
\begin{aligned}
\dot x(t)
&=e^{\lambda t}
\sum_{j=0}^{n-1}\frac{t^j}{j!}
(\lambda x_j+x_{j+1})\\
&=e^{\lambda t}
\sum_{j=0}^{n-1}\frac{t^j}{j!}
(Ax_j+Fd_j)\\
&=Ax(t)+Fd(t).
\end{aligned}
\]
Moreover,
\[
\begin{aligned}
Cx(t)+Gd(t)
&=e^{\lambda t}
\sum_{j=0}^{n-1}\frac{t^j}{j!}
(Cx_j+Gd_j)\\
&=0.
\end{aligned}
\]
Consequently, the functions in
\eqref{eq:chain-trajectory} satisfy
the zero-output system. If $\lambda$
or the coefficients are complex, their
real and imaginary parts provide
admissible real zero-output trajectories.

By SFD, the corresponding functional
outputs converge to zero. Consequently,
$
Lx(t)=e^{\lambda t}
\sum_{j=0}^{n-1}\frac{t^j}{j!}Lx_j
\longrightarrow0.
$
Since $\operatorname{Re}\lambda\geq0$,
an exponential-polynomial function of
this form can converge to zero only if
all its polynomial coefficients vanish.
Thus,
$
Lx_j=0,\qquad 0\leq j<n.
$
By the definition of $\mathcal F_n(\lambda)$,
these equalities imply
$\mathcal F_n(\lambda)\chi_n=0$.
Since $\chi_n$ was arbitrary, we conclude that
$
\ker\mathcal T_n(\lambda)
\subseteq\ker\mathcal F_n(\lambda),
$
Applying Lemma~3, we obtain the equivalent condition \eqref{eq:SFD-chain}.\\

\medskip
\noindent\emph{(2) $\Rightarrow$ (3).}
Assume \eqref{eq:SFD-chain} and $w>0$.
First, choose any sufficiently large real
$\lambda>0$ such that
$\lambda I_w-A_h$ is invertible.
For any $v\in\mathbb R^{\ell_h}$, set
$
\eta^\ast=(\lambda I_w-A_h)^{-1}B_hv
$
and define
$
x_0=V\eta^\ast,\qquad
d_0=K\eta^\ast+D_hv,
$
with $x_j=d_j=0$ for $1\leq j<n$
and $x_n=0$.
Using equations
\eqref{eq:SFD-reduced-matrices}, we obtain
\[
\begin{aligned}
A_\lambda x_0+Fd_0
&=V(A_h-\lambda I_w)\eta^\ast
  +VB_hv=0,\\
Cx_0+Gd_0&=0.
\end{aligned}
\]
It follows that
$\mathcal T_n(\lambda)\chi_n=0$.
By \eqref{eq:SFD-chain},
$\mathcal F_n(\lambda)\chi_n=0$,
and hence $Lx_0=0$.
Since $x_0=V(\lambda I_w-A_h)^{-1}B_hv$
and $C_h=LV$, we obtain
\[
C_h(\lambda I_w-A_h)^{-1}B_hv=0.
\]
Since the argument applies to every sufficiently
large real $\lambda$, the identity holds on an
interval. The resolvent expansion
$
(\lambda I_w-A_h)^{-1}
=\sum_{j=0}^{\infty}
\frac{A_h^j}{\lambda^{j+1}}
$
therefore gives
$
C_hA_h^jB_h=0,\qquad j\geq0.
$
In particular,
$C_h\mathcal R_h=0$, proving
\eqref{eq:SFD-reduced-condition1}.

Next, let $\lambda\in\sigma(A_h)$ with
$\operatorname{Re}\lambda\geq0$, and set
$m=m_h(\lambda)$.
For any
$v\in\ker(A_h-\lambda I_w)^m$, define
\[
\eta_j=(A_h-\lambda I_w)^jv,
\qquad 0\leq j\leq m.
\]
Since $\eta_m=0$, choose
\[
x_j=V\eta_j,\qquad d_j=K\eta_j,
\qquad 0\leq j<m,
\]
and extend these coefficients by zero
to length $n$.

By \eqref{eq:SFD-reduced-matrices},
\[
\begin{aligned}
A_\lambda x_j+Fd_j
&=V(A_h-\lambda I_w)\eta_j
=x_{j+1},\\
Cx_j+Gd_j&=0.
\end{aligned}
\]
Since $m\leq w\leq n$, this gives
a generalized chain of length $n$.
Condition \eqref{eq:SFD-chain} yields
$Lx_0=C_hv=0$.
As $v$ is arbitrary,
\[
\ker(A_h-\lambda I_w)^m
\subseteq\ker C_h,
\]
Applying Lemma~3, we obtain \eqref{eq:SFD-reduced-condition2}.

\medskip
\noindent\emph{(3) $\Rightarrow$ (1).}
Consider any admissible zero-output
trajectory. By \eqref{eq:SFD-reduced},
it admits the representation
\[
\dot\eta(t)=A_h\eta(t)+B_h\nu(t),
\qquad z(t)=C_h\eta(t).
\]
Condition \eqref{eq:SFD-reduced-condition1}
and the Cayley--Hamilton theorem give
\[
C_he^{A_ht}B_h=0,\qquad t\geq0.
\]
Moreover,
\eqref{eq:SFD-reduced-condition2}
implies that $C_h$ annihilates all
generalized eigenspaces of $A_h$
associated with eigenvalues satisfying
$\operatorname{Re}\lambda\geq0$.
Consequently,
\[
C_he^{A_ht}\longrightarrow0
\qquad (t\to\infty).
\]
On the other hand, we have
\[
\begin{aligned}
z(t)
&=C_he^{A_ht}\eta(0)\\
&\quad+\int_0^t
C_he^{A_h(t-\tau)}B_h\nu(\tau)\,d\tau\\
&=C_he^{A_ht}\eta(0)
\longrightarrow0.
\end{aligned}
\]
Thus, the system is strongly
functionally detectable.

Finally, if $w=0$, Lemma
\ref{lem:Kratz-SFD-completeness}
implies that every zero-output trajectory
satisfies $x(t)=0$.
Hence SFD holds, condition (2) follows
from the already established implication
$(1)\Rightarrow(2)$, and condition (3)
is automatically satisfied.
\end{proof}
\begin{remark}
The two conditions in item 3 are independent of the
admissible reduced matrices. For a fixed $V$, two choices
of $K$ give $A_{h,2}=A_{h,1}+B_hU_h$ for some $U_h$.
State feedback preserves the reachable subspace of
$(A_h,B_h)$; hence the first condition is unchanged.
When it holds, $C_hA_{h,2}^j=C_hA_{h,1}^j$ for all $j\geq0$,
so $C_he^{A_{h,2}t}=C_he^{A_{h,1}t}$.
Thus the spectral decay condition is unchanged as well.
Different bases of $\operatorname{Im}V$ and $\mathcal U_h$
only change coordinates.
\end{remark}
\color{black}

The following algorithm provides a step-by-step procedure for testing SFD.

\color{black}
\medskip\noindent\textbf{Algorithm 1}
\begin{enumerate}
\item Compute a basis matrix $\mathcal K_n$ of
$\ker\mathcal S_{(C,F,G,n)}$ and choose
\begin{equation}\label{eq:SFD-reduced-iteration}
\operatorname{Im}V=\operatorname{Im}(\Pi_0\mathcal K_n),
\end{equation}
where the columns of $V$ are linearly independent. If $\operatorname{rank}V=0$, declare SFD.
\color{black}

\item Otherwise, let $w$ be the number of
columns of $V$. Determine $K_h$, $A_h$,
$D_h$, and $B_h$ from
\begin{equation}\label{eq:SFD-reduced-defining}
\begin{aligned}
CV+GK_h&=0,& AV+FK_h&=VA_h,\\
GD_h&=0,& FD_h&=VB_h,\\
C_h&=LV,
\end{aligned}
\end{equation}
where $D_h\in\mathbb R^{d\times\ell_h}$,
$B_h\in\mathbb R^{w\times\ell_h}$,
$\ell_h=\dim\mathcal U_h$, and the columns of $D_h$ form a basis
of the unknown-input directions satisfying
$G\delta=0$ and
$F\delta\in\operatorname{Im}V$.
The reduced zero-output system is
\begin{equation}\label{eq:SFD-reduced-system}
\dot\eta (t) =A_h\eta (t) +B_h\nu (t) ,\qquad
z (t) =C_h\eta (t) .
\end{equation}

\item Form
\[
\mathcal R_h=
[\,B_h\ A_hB_h\ \cdots\
A_h^{w-1}B_h\,].
\]
Declare SFD if and only if
\[
C_h\mathcal R_h=0
\]
and
\[
\rank\begin{bmatrix}
(A_h-\lambda I_w)^{m_h(\lambda)}\\
C_h
\end{bmatrix}
=
\rank(A_h-\lambda I_w)^{m_h(\lambda)}
\]
for every $\lambda\in\sigma(A_h)$
with $\operatorname{Re}\lambda\geq0$.
\end{enumerate}
The following two examples illustrate the application of Algorithm 1.
\begin{example}
Consider the following simple system:
\[
\dot x(t)=ax(t)+d(t),\qquad
y(t)=x(t)-d(t)\quad \text{and} \quad z(t)=x(t).
\]
Here $A=a$, $F=C=L=1$, and $G=-1$.

\emph{Step 1.}
For $n=1$, the matrix defined in \eqref{eq:Kratz-general} is
\[
\mathcal S_{(C,F,G,1)}
=\begin{bmatrix}1&-1\end{bmatrix}.
\]
Since its kernel is spanned by
$(1,1)^\top$, we obtain $V=1$ and $w=1$.

\emph{Step 2.}
The equations
\[
CV+GK_h=0,\qquad
AV+FK_h=VA_h
\]
give $K_h=1$ and $A_h=a+1$.
Since $y(t)=0$ uniquely determines $d(t)=x(t)$,
the reduced zero-output system is autonomous.
Moreover, $C_h=LV=1$. Thus,
\[
\dot\eta(t)=(a+1)\eta(t),
\qquad z(t)=\eta(t).
\]

\emph{Step 3.}
Since $G=-1$ is nonsingular, the reachability
condition is automatically satisfied.
As $A_h=a+1$ and $C_h=1$, the generalized
eigenspace condition holds if and only if $a+1<0$.
Thus,
\[
\text{SFD}\quad\Longleftrightarrow\quad a<-1.
\]
For $a=-1/2$, $A$ is Hurwitz but SFD fails,
since $A_h=1/2$. This illustrates that
Algorithm~1 detects critical eigenvalues
outside $\sigma(A)$.
\end{example}
\begin{example}
Consider the system described by the following matrices:
$
A=\begin{bmatrix}
1&0&1\\
0&-1&0\\
0&0&-2
\end{bmatrix},$ 
$F=\begin{bmatrix}
0&1\\
0&0\\
0&0
\end{bmatrix},
$
$
C=\begin{bmatrix}
1&0&0\\
0&1&0
\end{bmatrix},$ 
$G=\begin{bmatrix}
1&0\\
0&0
\end{bmatrix},$ \text{and} $L=\begin{bmatrix}0&0&1\end{bmatrix}.
$

\emph{Step 1: Computation of $V$.}
For $n=3$, the Kratz construction
gives
\[
\mathcal S_{(C,F,G,3)}
=
\begin{bmatrix}
1&0&0&1&0&0&0&0&0\\
0&1&0&0&0&0&0&0&0\\
1&0&1&0&1&1&0&0&0\\
0&-1&0&0&0&0&0&0&0\\
1&0&-1&0&1&0&1&1&0\\
0&1&0&0&0&0&0&0&0
\end{bmatrix}.
\]
With
$\chi_3=\operatorname{col}(x_0,d_0,d_1,d_2)$,
the equation
$\mathcal S_{(C,F,G,3)}\chi_3=0$
reduces to
\[
\begin{aligned}
x_{0,2}&=0,\\
d_{0,1}&=-x_{0,1},\\
d_{1,1}&=-x_{0,1}-x_{0,3}-d_{0,2},\\
d_{2,1}&=-x_{0,1}+x_{0,3}
          -d_{0,2}-d_{1,2}.
\end{aligned}
\]
Consequently, the kernel has
dimension five. One basis matrix is
\[
\mathcal K_3=
\begin{bmatrix}
 1& 0& 0& 0&0\\
 0& 0& 0& 0&0\\
 0& 1& 0& 0&0\\
-1& 0& 0& 0&0\\
 0& 0& 1& 0&0\\
-1&-1&-1& 0&0\\
 0& 0& 0& 1&0\\
-1& 1&-1&-1&0\\
 0& 0& 0& 0&1
\end{bmatrix}.
\]
Thus,
\[
\operatorname{Im}(\Pi_0\mathcal K_3)
=\operatorname{span}\{e_1,e_3\}.
\]
We choose
\[
V=\begin{bmatrix}
1&0\\
0&0\\
0&1
\end{bmatrix},
\qquad w=2.
\]

\emph{Step 2: Reduced zero-output system.}
The defining equations
\[
\begin{aligned}
CV+GK_h&=0,\\
AV+FK_h&=VA_h,\\
GD_h&=0,\qquad FD_h=VB_h
\end{aligned}
\]
are satisfied by
\[
K_h=\begin{bmatrix}
-1&0\\
0&0
\end{bmatrix},\quad
A_h=\begin{bmatrix}
1&1\\
0&-2
\end{bmatrix},
\]
\[
D_h=\begin{bmatrix}0\\1\end{bmatrix},
\qquad
B_h=\begin{bmatrix}1\\0\end{bmatrix},
\qquad
C_h=LV=\begin{bmatrix}0&1\end{bmatrix}.
\]
The reduced zero-output system is
\[
\dot\eta(t)=A_h\eta(t)+B_h\nu(t),
\qquad z(t)=C_h\eta(t).
\]

\emph{Step 3: SFD tests.}
The reachability matrix is
\[
\mathcal R_h=[\,B_h\ \ A_hB_h\,]
=\begin{bmatrix}
1&1\\
0&0
\end{bmatrix}.
\]
Hence
\[
C_h\mathcal R_h=0.
\]
Moreover,
\[
\sigma(A_h)=\{1,-2\}.
\]
The only eigenvalue in the closed
right half-plane is $\lambda=1$,
with $m_h(1)=1$, and
\[
\ker(A_h-I_2)
=\operatorname{span}\{e_1\}
\subseteq\ker C_h.
\]
Both conditions of Algorithm~1
are satisfied. Therefore, the
system is SFD.

This conclusion can also be verified
directly. Since
\[
C_hA_h=-2C_h,\qquad C_hB_h=0,
\]
the reduced system gives
\[
\dot z(t)=-2z(t),
\qquad
z(t)=e^{-2t}z(0)\longrightarrow0.
\]
\end{example}
\begin{remark}
These examples illustrate that the eigenvalues
relevant to SFD need not belong to $\sigma(A)$
and that unstable zero-output modes do not
necessarily prevent SFD. Algorithm~1 accounts
for both situations through the reduced
zero-output dynamics.
\end{remark}
\subsection*{Particular cases of SFD}
We now derive several particular cases of the SFD.
from the preceding results. 
\begin{corol}
\label{cor:autonomous-FD}
For $F=G=0$, SFD reduces to functional
detectability, which holds if and only if
\begin{equation}\label{eq:FD-special-case}
\ker A_\lambda^{m_A(\lambda)}
\cap\ker\mathcal O(C,A,n)
\subseteq\ker L
\end{equation}
for every $\lambda\in\sigma(A)$ with
$\operatorname{Re}\lambda\geq0$, where
$A_\lambda=A-\lambda I_n$ and
$m_A(\lambda)$ is its algebraic multiplicity.
Equivalently, with
$
Q_\lambda=
\begin{bmatrix}
A_\lambda^{m_A(\lambda)}\\
\mathcal O(C,A,n)
\end{bmatrix},
$
\begin{equation}\label{eq:FD-special-case-rank}
\rank\begin{bmatrix}Q_\lambda\\L\end{bmatrix}
=\rank Q_\lambda.
\end{equation}
\end{corol}
%
\begin{proof}
For $F=G=0$,
$
\operatorname{Im}V=\ker\mathcal O(C,A,n),
\quad AV=VA_h,\quad C_h=LV.
$
The reduced dynamics are autonomous, so the
reachable-subspace condition is automatic.
Moreover, for every $k\geq1$,
$
V\ker(A_h-\lambda I_w)^k
=\ker\mathcal O(C,A,n)\cap\ker A_\lambda^k.
$
Since $A_h$ represents the restriction of
$A$ to $\ker\mathcal O(C,A,n)$, we have
$\sigma(A_h)\subseteq\sigma(A)$ and
\[
V\ker(A_h-\lambda I_w)^{m_h(\lambda)}
=\ker\mathcal O(C,A,n)
\cap\ker A_\lambda^{m_A(\lambda)}.
\]
Thus, using $C_h=LV$, the reduced SFD
criterion gives \eqref{eq:FD-special-case}.
For $\lambda\in\sigma(A)\setminus\sigma(A_h)$,
the intersection is zero. Finally,
Lemma~\ref{lem2} yields
\eqref{eq:FD-special-case-rank}.
\end{proof}
\begin{corol}[Classical state detectability]
\label{cor:autonomous-Hautus}
If $F=G=0$ and $L=I_n$, SFD reduces to
\begin{equation}\label{eq:Hautus-explicit}
\rank\begin{bmatrix}\lambda I_n-A\\C\end{bmatrix}=n,
\qquad\operatorname{Re}\lambda\geq0.
\end{equation}
\end{corol}
\begin{proof}
With $L=I_n$, condition \eqref{eq:FD-special-case}
requires
$\ker A_\lambda^n\cap\mathcal N=\{0\}$
for every nonstable eigenvalue. If a nonzero vector
belonged to this intersection, the restriction of
$A_\lambda$ to that nonzero invariant subspace
would be nilpotent, and would therefore have a nonzero
vector $v$ satisfying $A_\lambda v=0$ and $Cv=0$.
Conversely, any such eigenvector belongs to the
intersection. Therefore the intersection is zero
exactly when
$\ker(\lambda I_n-A)\cap\ker C=\{0\}$,
which is \eqref{eq:Hautus-explicit}.
\end{proof}
\subsection{Strong* functional detectability}
We now turn to S*FD, defined as follows.
\begin{definition}
\label{def:sstar-original}
System~\eqref{sys-kratz-md} is strongly-star
functionally detectable (S$^*$FD) with respect to $z(t)=Lx(t)$ if, for every
admissible initial state and unknown input,
\begin{equation}\label{eq:sstar-functional-definition}
y(t)\longrightarrow0
\quad\Longrightarrow\quad
Lx(t)\longrightarrow0,
\qquad t\to\infty.
\end{equation}
\end{definition}
{\color{black}
Although SFD guarantees that $Lx(t)\to0$ whenever
$y(t)\equiv0$, it does not necessarily guarantee the same
convergence when $y(t)\to0$. Unknown inputs may produce
state variations whose effect on the measured output tends
to zero without a corresponding decay of the prescribed
functional. S$^*$FD therefore requires an additional
algebraic condition accounting for these input directions.

To formulate the additional condition for S*FD,
consider the sequence of subspaces generated by
\[
\mathcal S_0=\{0\},\qquad
\mathcal S_{j+1}
=\{Av+F\delta:\,
v\in\mathcal S_j,\ Cv+G\delta=0\}.
\]
These subspaces form an increasing sequence
that stabilizes by step $n$.

For computation, choose a full-column-rank
matrix $V_1$ such that
\[
\operatorname{Im}V_1=F\ker G,
\]
and recursively determine $V_{j+1}$ from
\begin{equation}\label{eq:S-infinity-recursion}
\operatorname{Im}V_{j+1}
=[\,AV_j\ \ F\,]\ker[\,CV_j\ \ G\,],
\quad j=1,\ldots,n-1.
\end{equation}
Let $V_\infty$ be a full-column-rank
matrix satisfying
\[
\operatorname{Im}V_\infty=\mathcal S_n.
\]

The additional condition concerns state directions
in $\mathcal S_n$ compatible with the current
zero-output constraint:
\begin{equation}\label{eq:S-infinity-property}
v\in\operatorname{Im}V_\infty,\quad
Cv+G\delta=0
\quad\Longrightarrow\quad Lv=0.
\end{equation}
To express it as a rank condition, let
$G_\perp$ be a full-row-rank matrix whose
rows form a basis of $\ker G^\top$.
Then
\[
G_\perp G=0,\qquad
\ker G_\perp=\operatorname{Im}G.
\]
Since $Cv+G\delta=0$ is solvable in
$\delta$ if and only if $G_\perp Cv=0$,
condition \eqref{eq:S-infinity-property}
is equivalent to
\[
\ker(G_\perp CV_\infty)
\subseteq\ker(LV_\infty).
\]
By Lemma~3, this yields
\begin{equation}\label{eq:S-infinity-rank}
\operatorname{rank}
\begin{bmatrix}
G_\perp CV_\infty\\
LV_\infty
\end{bmatrix}
=
\operatorname{rank}(G_\perp CV_\infty).
\end{equation}
Together with SFD, this condition is used
in the following characterization of S*FD.
\begin{thm}[S$^*$FD]
\label{thm:general-hurwitz-existence}
For system~\eqref{sys-kratz-md}, the following statements are equivalent:
\begin{enumerate}
\item The system satisfies S$^*$FD:
$y(t)\to0$ implies $Lx(t)\to0$.
\item The system is SFD and
satisfies~\eqref{eq:S-infinity-rank}.
\item There exist rational matrices $M(s)$ and $\mathcal H(s)$
with no poles in the closed right half-plane, with $\mathcal H(s)$
proper, such that
\begin{equation}\label{eq:sstar-stable-bezout}
M(s)\mathcal P(s)+\mathcal H(s)\mathcal Q=\mathcal L_0,
\end{equation}
where
\[
\begin{aligned}
\mathcal P(s)&=[\,sI_n-A\quad-F\,],\\
\mathcal Q&=[\,C\quad G\,].
\end{aligned}
\]
\end{enumerate}
\end{thm}
\begin{proof}
The equivalence $(1)\Leftrightarrow(3)$
follows from \cite[Theorem~3.2]{28}.

\emph{(1) $\Rightarrow$ (2).}
S$^*$FD implies SFD.
Moreover, by (3), there exist stable
rational matrices $M(s)$ and
$\mathcal H(s)$ satisfying
\eqref{eq:sstar-stable-bezout},
with $\mathcal H(s)$ proper.
The state columns give
\[
M(s)=[L-\mathcal H(s)C](sI_n-A)^{-1},
\]
so $M(s)$ is strictly proper.

Post-multiplying \eqref{eq:sstar-stable-bezout}
by $\operatorname{col}(p(s),q(s))$ yields
\[
\begin{aligned}
Lp(s)={}&M(s)\big[(sI_n-A)p(s)-Fq(s)\big]\\
&+\mathcal H(s)\big[Cp(s)+Gq(s)\big].
\end{aligned}
\]
Since $M(s)$ and $\mathcal H(s)$ are proper,
$Lp(s)$ is proper whenever the two
expressions in brackets are proper.

To establish \eqref{eq:S-infinity-property},
let $v\in\mathcal S_n$ satisfy
$Cv+G\delta=0$ for some $\delta$.
By the recursion defining $\mathcal S_n$,
there exist $w_0,\ldots,w_n$ and
$d_0,\ldots,d_{n-1}$ satisfying
\[
\begin{aligned}
&w_0=0,\qquad w_n=v,\\
&w_{j+1}=Aw_j+Fd_j,\\
&Cw_j+Gd_j=0,\qquad 0\leq j<n.
\end{aligned}
\]
Set $a=Av+F\delta$ and define
\[
\begin{aligned}
p(s)&=a+\sum_{j=1}^{n}s^jw_{n+1-j},\\
q(s)&=s\delta+
\sum_{j=2}^{n+1}s^jd_{n+1-j}.
\end{aligned}
\]
The state recursion and zero-output
relations cancel all positive-degree
terms, giving
\[
\begin{aligned}
(sI_n-A)p(s)-Fq(s)&=-Aa,\\
Cp(s)+Gq(s)&=Ca.
\end{aligned}
\]
Then,
\[
Lp(s)=-M(s)Aa+\mathcal H(s)Ca
\]
is proper. Since $Lp(s)$ is polynomial,
its coefficient of $s$ must vanish.
Thus $Lv=0$, proving
\eqref{eq:S-infinity-property}.

Finally, since
$\operatorname{Im}V_\infty=\mathcal S_n$
and $\ker G_\perp=\operatorname{Im}G$,
we obtain
\[
\ker(G_\perp CV_\infty)
\subseteq\ker(LV_\infty),
\]
which, by Lemma~3, is equivalent to
\eqref{eq:S-infinity-rank}.

\medskip
\noindent\emph{(2) $\Rightarrow$ (3).}
Assume SFD and \eqref{eq:S-infinity-rank}.
The rank condition is equivalent to
\eqref{eq:S-infinity-property}.
The stable rational solvability criterion and the
properness criterion of Hautus~\cite[Theorems~3.8
and~2.1]{28}, applied to  equation
\eqref{eq:sstar-stable-bezout}, yield a stable
rational solution for which $\mathcal H(s)$ is proper.
This equation also  give
\[
M(s)=[L-\mathcal H(s)C](sI_n-A)^{-1},
\]
so $M(s)$ is strictly proper as well.
This proves (3).
\end{proof}
\subsection*{Particular cases of S*FD}
\begin{corollary}
\label{cor:hautus-infinity-recovery}
Let $L=I_n$. Let the columns of $K_G$ form a basis
of $\ker G$, and let $G_\perp$ be a matrix satisfying
$\ker G_\perp=\operatorname{Im}G$.
Then the following conditions are equivalent:
\begin{align}
&\operatorname{rank}(G_\perp C V_\infty)
 =\operatorname{rank}V_\infty,
\label{eq:hautus-from-recursion}\\
&\operatorname{rank}(G_\perp CFK_G)
 =\operatorname{rank}(FK_G),
\label{eq:hautus-first-step}\\
&\operatorname{rank}
\begin{bmatrix}
CF&G\\
G&0
\end{bmatrix}
=
\operatorname{rank}G+
\operatorname{rank}
\begin{bmatrix}
F\\G
\end{bmatrix}.
\label{eq:hautus-original-rank}
\end{align}
Consequently, for $L=I_n$, the algebraic rank condition
reduces exactly to the classical
Hautus condition~\cite{28}. No full-column-rank
assumption on $\begin{bmatrix}F^\top&G^\top
\end{bmatrix}^\top$ is required.
\end{corollary}
\begin{proof}
By the definition of the recursive spaces,
\[
\mathcal S_1=F\ker G=\operatorname{Im}(FK_G).
\]
Since $\mathcal S_1\subseteq\mathcal S_n$,
\eqref{eq:hautus-from-recursion} implies
\eqref{eq:hautus-first-step}.

Conversely, suppose that
\eqref{eq:hautus-first-step} holds.
Then $G_\perp C$ is injective on
$\mathcal S_1$. Thus, if
$v\in\mathcal S_1$ satisfies
$Cv+G\delta=0$, we have
\[
G_\perp Cv=0,
\]
and therefore $v=0$. It follows that
\[
\mathcal S_2
=\{F\delta:G\delta=0\}
=\mathcal S_1.
\]
The recursion then gives
$\mathcal S_n=\mathcal S_1$, proving
\eqref{eq:hautus-from-recursion}.

It remains to establish the equivalence
with \eqref{eq:hautus-original-rank}.
Let $g=\operatorname{rank}G$.
Choose a matrix $T_G$ such that
$[T_G\ K_G]$ is nonsingular and
$\operatorname{rank}(GT_G)=g$.
Multiplication by this matrix in the
appropriate block columns gives
\[
\operatorname{rank}
\begin{bmatrix}F\\G\end{bmatrix}
=g+\operatorname{rank}(FK_G).
\]

Similarly, applying the same column
transformation to the first block column
of $\begin{bmatrix}CF&G\\G&0\end{bmatrix}$
and eliminating the independent columns
associated with $GT_G$ and $G$ gives
\[
\operatorname{rank}
\begin{bmatrix}
CF&G\\G&0
\end{bmatrix}
=
2g+\operatorname{rank}(G_\perp CFK_G).
\]
Then \eqref{eq:hautus-original-rank}
is equivalent to
\[
\operatorname{rank}(G_\perp CFK_G)
=\operatorname{rank}(FK_G),
\]
which is \eqref{eq:hautus-first-step}.
\end{proof}
\medskip

\noindent\textbf{Algorithm 2.}
\begin{enumerate}
\item Apply Algorithm~1. If the system is not
SFD, it is not S$^*$FD, and the procedure
terminates.

\item Choose a full-column-rank matrix $V_1$
such that
\[
\operatorname{Im}V_1=F\ker G.
\]
For $j=1,\ldots,n-1$, compute
\begin{equation}\label{eq:strongstar-direct-iteration}
\operatorname{Im}V_{j+1}
=
[\,AV_j\ \ F\,]\ker[\,CV_j\ \ G\,],
\end{equation}
where $V_{j+1}$ is chosen with
linearly independent columns.

Stop when
$\operatorname{rank}V_{j+1}
=\operatorname{rank}V_j$,
or when $j=n-1$.
Denote the final matrix by $V_\infty$.

\item Choose a full-row-rank matrix $G_\perp$
whose rows span $\ker G^\top$.
Compute
\[
R_0=G_\perp CV_\infty,\qquad
R_1=
\begin{bmatrix}
R_0\\LV_\infty
\end{bmatrix}.
\]
The system is S$^*$FD if and only if
\begin{equation}\label{eq:sstar-algorithm-rank}
\operatorname{rank}R_1
=\operatorname{rank}R_0.
\end{equation}
\end{enumerate}
The following example illustrates Algorithm $2$.
\begin{example}
Consider the system described by the following matrices:
$
A=\begin{bmatrix}
-1&1\\0&-2
\end{bmatrix},$ $
F=\begin{bmatrix}0\\1\end{bmatrix},$ $C=\begin{bmatrix}1&0\end{bmatrix},$
 $G=0,$ and 
$L=\begin{bmatrix}0&1\end{bmatrix}.$

\emph{Step 1: SFD test (Algorithm~1).}
For $n=2$, the matrix associated with
the Kratz construction is
\[
\mathcal S_{(C,F,G,2)}
=
\begin{bmatrix}
1&0&0&0\\
-1&1&0&0
\end{bmatrix}.
\]
Let $\mathcal K_2$ be a basis matrix
of its kernel. Since
\[
\mathcal K_2=
\begin{bmatrix}
0&0\\
0&0\\
1&0\\
0&1
\end{bmatrix},
\qquad
\Pi_0=\begin{bmatrix}I_2&0_{2\times2}\end{bmatrix},
\]
we obtain
\[
\operatorname{rank}(\Pi_0\mathcal K_2)=0.
\]
Thus, the matrix $V$ computed in
Step~1 of Algorithm~1 has no columns.
According to Algorithm~1, the system
is SFD, and no reduced-system
calculation is required.

\emph{Step 2: Computation of $V_\infty$.}
Since $\ker G=\mathbb R$, choose
\[
V_1=F=\begin{bmatrix}0\\1\end{bmatrix}.
\]
Then
\[
[\,CV_1\ \ G\,]
=\begin{bmatrix}0&0\end{bmatrix},
\]
so its kernel is $\mathbb R^2$.
Consequently,
\[
\begin{aligned}
\operatorname{Im}V_2
&=[\,AV_1\ \ F\,]
\ker[\,CV_1\ \ G\,]\\
&=\operatorname{Im}
\begin{bmatrix}
1&0\\-2&1
\end{bmatrix}
=\mathbb R^2.
\end{aligned}
\]
Since $n=2$, the iteration terminates
with
\[
V_\infty=I_2.
\]

\emph{Step 3: Rank test.}
For $G=0$, choose $G_\perp=1$.
We obtain
\[
R_0=G_\perp CV_\infty
=\begin{bmatrix}1&0\end{bmatrix},
\]
and
\[
R_1=
\begin{bmatrix}
G_\perp CV_\infty\\
LV_\infty
\end{bmatrix}
=
\begin{bmatrix}
1&0\\0&1
\end{bmatrix}.
\]
Therefore,
\[
\operatorname{rank}R_1=2
>1=\operatorname{rank}R_0.
\]
The rank condition fails, and
Algorithm~2 shows that the system
is not S$^*$FD.

To illustrate this failure,
the first state equation gives
\[
z(t)=\dot y(t)+y(t).
\]
Choose
\[
y(t)=\frac{\sin(t^2)}{t+1}.
\]
Although $y(t)\to0$, we have
\[
\begin{aligned}
z(t)={}&
\frac{2t\cos(t^2)}{t+1}
-\frac{\sin(t^2)}{(t+1)^2}\\
&+\frac{\sin(t^2)}{t+1},
\end{aligned}
\]
which does not converge to zero.
This trajectory is admissible with
$x_1(t)=y(t)$,
$x_2(t)=\dot y(t)+y(t)$, and
$d(t)=\dot x_2(t)+2x_2(t)$.

Thus, the system is SFD but
not S$^*$FD.
\end{example}
\section{Functional Unknown-Input Observer Synthesis}
In this section, we investigate the existence
and design of stable functional observers for
linear systems with unknown inputs.
We first introduce the observer structure and
derive the corresponding existence conditions.
We then develop two constructive design procedures.
\subsection{Observer Structure and Existence Conditions}
\label{subsec:observer-existence}
Since $u(t)$ is a known input, the SFO, SFD,
and S$^*$FD properties of system~(1) are
equivalent to those of system~(12).
In the following, the functional observer
is designed for system~(1), explicitly
taking into account the known input $u(t)$.
We now consider the design of a stable functional observer for system
\eqref{sys-kratz} of the form
\begin{align}
\dot{\xi}&=N\xi+Jy+B_{\mathrm{o}}u,
\label{eq20a}\\
\hat z&=P\xi+J_1y,
\label{eq20b}
\end{align}
where $N\in\mathbb R^{q\times q}$,
$J\in\mathbb R^{q\times p}$,
$B_{\mathrm{o}}\in\mathbb R^{q\times m}$,
$P\in\mathbb R^{r\times q}$, and
$J_1\in\mathbb R^{r\times p}$.
The pair $(N,P)$ is assumed observable.

For $T\in\mathbb R^{q\times n}$, define
$\varepsilon=\xi-Tx$. From the plant and
observer equations,
\begin{align}
\dot{\varepsilon}
={}&N\varepsilon+(NT-TA+JC)x
\nonumber\\
&+(B_{\mathrm{o}}-TB)u+(JG-TF)d,
\label{eq21}\\
\hat z-z
={}&P\varepsilon+(PT+J_1C-L)x+J_1Gd.
\label{eq22}
\end{align}
Thus, the estimation error is independent
of the state and unknown input if
\begin{subequations}\label{eq23}
\begin{align}
NT-TA+JC&=0, \label{eq23-a}\\
B_{\mathrm{o}}&=TB, \label{eq23-b}\\
JG-TF&=0, \label{eq23-c}\\
J_1G&=0, \label{eq23-d}\\
PT+J_1C&=L. \label{eq23-e}
\end{align}
\end{subequations}
Under these conditions,
\[
\dot{\varepsilon}=N\varepsilon,\qquad
\hat z-z=Pe^{Nt}\varepsilon(0).
\]
Since $(N,P)$ is observable, the output
estimation error converges to zero for
every initial condition if and only if
$N$ is Hurwitz. Moreover, once $T$ is
known, $B_{\mathrm{o}}=TB$ follows directly.
The following theorem concerns the existence of a functional observer in presence of unknown inputs.
\begin{thm}
\label{stable-realization}
There exist matrices $T,J,J_1,P,B_{\mathrm{o}}$
and a Hurwitz matrix $N$, with $(N,P)$
observable, satisfying \eqref{eq23}
if and only if system \eqref{sys-kratz}
is strongly* functionally detectable.
\end{thm}

\begin{proof}
\emph{Necessity.}
Suppose that \eqref{eq23} holds with $N$
Hurwitz. Then
$\dot{\varepsilon}=N\varepsilon$ and
$\hat z(t)-z(t)=P\varepsilon(t)$, so
$\hat z(t)-z(t)\to0$.

For $u=0$, if $y(t)\to0$, the stable
observer equation
$\dot{\xi}=N\xi+Jy$ implies
$\xi(t)\to0$. Consequently,
$\hat z(t)=P\xi(t)+J_1y(t)\to0$,
and therefore $z(t)\to0$.
Thus the system is strongly*
functionally detectable.

\emph{Sufficiency.}
Suppose that the system is strongly*
functionally detectable. By the preceding
rational characterization, there exist
stable rational matrices $M(s)$ and $\mathcal H(s)$,
with $M(s)$ strictly proper and $\mathcal H(s)$
proper, such that
\begin{align}
M(s)(sI_n-A)+\mathcal H(s)C&=L,
\label{eq:observer-rational-1}\\
M(s)F&=\mathcal H(s)G.
\label{eq:observer-rational-2}
\end{align}
Set $J_1=\mathcal H(\infty)$ and
$\mathcal H_0(s)=\mathcal H(s)-J_1$. The rational matrix
$[M(s)\ \ \mathcal H_0(s)]$ is stable and
strictly proper. Choose a minimal
realization
\begin{equation}
[M(s)\ \ \mathcal H_0(s)]
=P(sI_q-N)^{-1}[T\ \ J],
\label{eq:observer-common-realization}
\end{equation}
where $N$ is Hurwitz and $(N,P)$
is observable.

Substituting this realization into
\eqref{eq:observer-rational-1} and using
\[
T(sI_n-A)+JC
=(sI_q-N)T+(NT-TA+JC)
\]
gives
\[
L=PT+J_1C+
P(sI_q-N)^{-1}(NT-TA+JC).
\]
Letting $s\to\infty$ yields
$L=PT+J_1C$. Consequently,
\[
P(sI_q-N)^{-1}(NT-TA+JC)=0.
\]
Observability of $(N,P)$ implies
$NT-TA+JC=0$.

Similarly, \eqref{eq:observer-rational-2}
gives
\[
P(sI_q-N)^{-1}(TF-JG)=J_1G.
\]
Letting $s\to\infty$ yields $J_1G=0$,
and observability gives $TF=JG$.
Finally, choosing $B_{\mathrm{o}}=TB$
establishes all conditions in
\eqref{eq23}.
\end{proof}
At this stage, the only structural assumption required by the observer
design is that the pair \((N,P)\) be observable. In particular, no
assumption is made on its observability indices.
Consequently, the classical Brunovský observable canonical form cannot
be employed directly, since its construction requires the prior
knowledge of the observability indices
$
(\nu_1,\ldots,\nu_r),
$
which are generally unknown.
As in \cite{16} we therefore consider the special case in
which all observability indices are equal. As can be seen in the sequel , this leads to
a particularly simple block companion realization that is well suited
for the proposed observer design.
\begin{lemma}\label{lem:balanced-brunovsky}
Let
$
N\in\mathbb R^{kr\times kr},
$
$P\in\mathbb R^{r\times kr},
$
and assume that the pair \((N,P)\) is observable with observability
indices
$
\nu_1=\nu_2=\cdots=\nu_r=k.
$
Equivalently, the observable pair \((N,P)\) consists of \(r\)
Brunovský observable chains, all having the same length \(k\). Then
$
kr=\nu_1+\cdots+\nu_r
$
and there exists a nonsingular matrix \(S\) such that
$
\bar N=SNS^{-1},
\qquad
\bar P=PS^{-1},
$
where
\small
\begin{equation}
\label{eq:NPcanonical}
\begin{aligned}
\bar N&=
\begin{pmatrix}
0&I_r&0&\cdots&0\\
0&0&I_r&\cdots&0\\
\vdots&\vdots&\vdots&\ddots&\vdots\\
0&0&0&\cdots&I_r\\
N_1&N_2&\cdots&N_{k-1}&N_k
\end{pmatrix},\\[2mm]
\bar P&=
\begin{pmatrix}
I_r&0&\cdots&0
\end{pmatrix}.
\end{aligned}
\end{equation}
\normalsize
for uniquely determined matrices
$
N_i\in\mathbb R^{r\times r},
\quad
i=1,\ldots,k.
$
\end{lemma}
\begin{proof}
Set
$
S=\operatorname{col}(P,PN,\ldots,PN^{k-1}).
$
Since the pair $(N,P)$ has equal observability
indices $k$, the matrix $S$ is square and
nonsingular. Consequently, there exist unique
matrices $N_i\in\mathbb R^{r\times r}$ such that
$
PN^k=\sum_{i=1}^{k}N_iPN^{i-1}.
$
Indeed, the block rows of $S$ form a basis,
so $PN^k$ has a unique representation in
terms of these block rows.
Multiplying $S$ by $N$ and using this identity
gives
$
SN=
\begin{bmatrix}
0&I_r&\cdots&0\\
\vdots&&\ddots&\vdots\\
0&0&\cdots&I_r\\
N_1&N_2&\cdots&N_k
\end{bmatrix}S.
$
Thus $\bar N=SNS^{-1}$ has the required
block companion form. Moreover, since $P$
is the first block row of $S$,
\[
\bar P=PS^{-1}=[I_r\ \ 0\ \cdots\ \ 0].
\]
\end{proof}
We now derive the design equations directly from the balanced
realization~\eqref{eq:NPcanonical}.  In this realization the
observer has $k$ block rows of size $r$ and order $kr$.
For simplicity we omit the bars on $N$ and $P$ and write
\begin{equation}\label{eq:TJ-partition}
 T=\operatorname{col}(T_1,\ldots,T_k),\qquad
 J=\operatorname{col}(J_2,\ldots,J_{k+1}).
\end{equation}
From  \eqref{eq23-e} we deduce $T_1=L-J_1C$.
The first $k-1$ block rows of  equation
\eqref{eq23-a} give $T_{i+1}=T_iA-J_{i+1}C$.
Consequently, by induction,
\begin{equation}\label{eq24c}
 T_i=LA^{i-1}-\sum_{j=1}^{i}J_jCA^{i-j},
 \quad 1\leq i\leq k.
\end{equation}
The last block row is
\begin{equation}\label{eq24d}
 T_kA=\sum_{i=1}^kN_iT_i+J_{k+1}C.
\end{equation}
Substituting \eqref{eq24c} into \eqref{eq24d}  yields
\begin{equation}\label{eq--33}
 LA^k=\sum_{i=1}^kN_iLA^{i-1}
       +\sum_{i=1}^{k+1}\varphi_iCA^{i-1}.
\end{equation}
where the matrices $\varphi_i$ are defined by
\begin{equation}\label{eq--38}
\begin{aligned}
\varphi_{k+2-i}={}&J_i
 -\sum_{j=1}^{i-1}N_{k-j+1}J_{i-j},\\[-1mm]
&1\leq i\leq k+1.
\end{aligned}
\end{equation}
From this equation we can see that when $N_i$ and $\varphi_i$ are known we can deduce matrices $J_i$. On the other hand from equations \eqref{eq23-c} and \eqref{eq23-d}
we have
\begin{equation}\label{eq--40}
 J_1G=0,
\end{equation}
\begin{equation}\label{eq--41}
 T_iF=J_{i+1}G,\qquad 1\leq i\leq k.
\end{equation}
Using \eqref{eq--38}, \eqref{eq24c},
\eqref{eq--40}, and \eqref{eq--41},
we obtain
\begin{equation}\label{eq:LAFrec}
\begin{aligned}
LA^{i-1}F
={}&\sum_{j=1}^{i-1}
N_{k-j+1}LA^{i-j-1}F\\
&+\sum_{j=1}^{i}
\varphi_{k+2-j}CA^{i-j}F\\
&+\varphi_{k-i+1}G,
\quad 1\le i\le k.
\end{aligned}
\end{equation}
Moreover, \eqref{eq--38} and \eqref{eq--40}
imply
$
\varphi_{k+1}G=0.
$
Define
\begin{equation}\label{eq:design-YW}
\begin{aligned}
Y_k&=
\begin{bmatrix}
\mathcal S_{(L,F,0,k)}&0_{kr\times d}
\end{bmatrix},\\
W_k&=\mathcal S_{(C,F,G,k+1)},\\
\Sigma_k&=\begin{bmatrix}Y_k\\W_k\end{bmatrix}.
\end{aligned}
\end{equation}
Here, $\mathcal S_{(L,F,0,k)}$ denotes
$\mathcal S_{(C,F,G,k)}$ with $C=L$ and $G=0$.
Furthermore, define
\begin{equation}\label{eq:design-Theta}
\Theta_{(L,F,k)}
=\begin{bmatrix}
LA^k&LA^{k-1}F&\cdots&LF&0_{r\times d}
\end{bmatrix}.
\end{equation}
Equations \eqref{eq--33}, \eqref{eq:LAFrec},
and $\varphi_{k+1}G=0$ can therefore be
collected into the single matrix equation
\begin{equation}\label{eqMMD}
\underbrace{\begin{bmatrix}
N_1&\cdots&N_k&
\varphi_1&\cdots&\varphi_{k+1}
\end{bmatrix}}_{\Omega_k}
\Sigma_k=\Theta_{(L,F,k)}.
\end{equation}
Now, define the $kr\times
[n+(k+1)d]$ matrix
\begin{equation}\label{eq:RLF}
\begin{aligned}
\mathcal R_{(L,F,\lambda,k)}
&=\operatorname{col}(R_k(\lambda),\ldots,R_1(\lambda)),\\
\text{with} \quad
R_j(\lambda)
&=\begin{bmatrix}
\lambda^jL-LA^j&-LA^{j-1}F&\cdots&-LF&0
\end{bmatrix},
\end{aligned}
\end{equation}
The following lemma establishes necessary conditions
for S$^*$FD in terms of
the solvability of \eqref{eqMMD} and a
frequency-dependent rank equality.
\begin{lemma}
\label{lem:strongstar-consistency}
If system~\eqref{sys-kratz} is strong$^*$
functionally detectable, then there exists an
integer $k\geq1$ such that \eqref{eqMMD}
admits a solution, or equivalently,
\begin{equation}\label{S_46}
\rank\begin{bmatrix}
\Theta_{(L,F,k)}\\
\Sigma_k
\end{bmatrix}
=\rank\Sigma_k.
\end{equation}
Moreover, for the same integer $k$ and every
$\lambda\in\mathbb C$ with
$\operatorname{Re}\lambda\geq0$,
\begin{equation}\label{S_47}
\rank\begin{bmatrix}
\mathcal R_{(L,F,\lambda,k)}\\
\mathcal S_{(C,F,G,k+1)}
\end{bmatrix}
=\rank\Sigma_k.
\end{equation}
\end{lemma}
\begin{proof}
By Theorem~4, S$^*$FD guarantees the existence
of a stable functional observer
$(T,N,J,P,J_1)$ satisfying
\begin{equation}\label{eq:observer-toeplitz-identities}
\begin{aligned}
TA-NT&=JC, & TF&=JG,\\
L&=PT+J_1C, & J_1G&=0,
\end{aligned}
\end{equation}
where $N\in\mathbb R^{s\times s}$ is Hurwitz.
Let
$
p_N(\mu)=\mu^s+\sum_{j=0}^{s-1}a_j\mu^j
$
be the characteristic polynomial of $N$.
By the Cayley--Hamilton theorem,
\begin{equation}\label{eq:CH-observer-N}
N^s+\sum_{j=0}^{s-1}a_jN^j=0.
\end{equation}
We choose $k=s$.
Let
$
\chi_{k+1}
=\operatorname{col}(x_0,d_0,\ldots,d_k)
$
and define the algebraic coefficients by
$
x_{j+1}=Ax_j+Fd_j,
\quad 0\leq j<k.
$
By construction,
\[
\begin{aligned}
Y_k\chi_{k+1}
 &=\operatorname{col}(Lx_0,\ldots,Lx_{k-1}),\\
W_k\chi_{k+1}
 &=\operatorname{col}(Cx_0+Gd_0,\ldots,
                     Cx_k+Gd_k),\\
\Theta_{(L,F,k)}\chi_{k+1}
 &=Lx_k.
\end{aligned}
\]
Suppose that $W_k\chi_{k+1}=0$.
The observer matching equations imply
\[
\begin{aligned}
Tx_{j+1}
 &=NTx_j+J(Cx_j+Gd_j)=NTx_j,\\
Lx_j
 &=PTx_j+J_1Cx_j=PTx_j,
\end{aligned}
\]
where the second equality follows from
$Cx_j=-Gd_j$ and $J_1G=0$.
Consequently,
\begin{equation}\label{eq:observer-chain-recurrence}
Lx_j=PN^jTx_0,\qquad 0\leq j\leq k.
\end{equation}
Applying \eqref{eq:CH-observer-N} gives
$
Lx_k+\sum_{j=0}^{k-1}a_jLx_j=0.
$
Thus, if $\Sigma_k\chi_{k+1}=0$,
then $Lx_j=0$ for $0\leq j<k$,
and therefore $Lx_k=0$. Thus,
$
\ker\Sigma_k
\subseteq\ker\Theta_{(L,F,k)}.
$
Lemma 3 proves \eqref{S_46} and,
equivalently, the solvability of
\eqref{eqMMD}.
We now prove \eqref{S_47}. Let
$\lambda\in\mathbb C$ satisfy
$\operatorname{Re}\lambda\geq0$ and put
$v=Tx_0$.
For every $\chi_{k+1}\in\ker W_k$,
the definition of $R_j(\lambda)$ gives
\[
\begin{aligned}
R_j(\lambda)\chi_{k+1}
&=\lambda^jLx_0-LA^jx_0
  -\sum_{i=0}^{j-1}LA^{j-1-i}Fd_i\\
&=\lambda^jLx_0-Lx_j\\
&=P(\lambda^jI_s-N^j)v,
\end{aligned}
\]
for $1\leq j\leq k$, where the last
equality follows from
\eqref{eq:observer-chain-recurrence}.
Using the factorization
$
\lambda^jI_s-N^j
=\sum_{i=0}^{j-1}
\lambda^{j-1-i}N^i(\lambda I_s-N),
$
and setting $w=(\lambda I_s-N)v$,
the equations
$R_j(\lambda)\chi_{k+1}=0$,
$1\leq j\leq k$, become
$
P\left(\sum_{i=0}^{j-1}
\lambda^{j-1-i}N^i\right)w=0.
$
For $j=1$, this gives $Pw=0$;
for $j=2$, it then gives $PNw=0$.
Proceeding inductively, we obtain
$
\mathcal R_{(L,F,\lambda,k)}
\chi_{k+1}=0
\iff
PN^jw=0,\quad 0\leq j<k.
$
Define
$
\mathcal K_k
=\bigcap_{j=0}^{k-1}\ker(PN^j).
$
Since $k=s$, the Cayley--Hamilton
theorem implies that $\mathcal K_k$
is $N$-invariant. Indeed, if
$v\in\mathcal K_k$, then
$PN^j(Nv)=0$ for $0\leq j<k-1$,
while $PN^kv=0$ follows from
the Cayley--Hamilton identity.

Since $N$ is Hurwitz and
$\operatorname{Re}\lambda\geq0$,
the restriction of $\lambda I_s-N$
to $\mathcal K_k$ is invertible.
Consequently,
\[
(\lambda I_s-N)v\in\mathcal K_k
\iff v\in\mathcal K_k.
\]
Therefore, whenever
$W_k\chi_{k+1}=0$,
\[
\begin{aligned}
\mathcal R_{(L,F,\lambda,k)}
\chi_{k+1}=0
&\iff v\in\mathcal K_k\\
&\iff PN^jv=0,\quad 0\leq j<k\\
&\iff Y_k\chi_{k+1}=0.
\end{aligned}
\]
It follows that
\[
\ker\begin{bmatrix}
\mathcal R_{(L,F,\lambda,k)}\\
W_k
\end{bmatrix}
=
\ker\begin{bmatrix}
Y_k\\W_k
\end{bmatrix}
=\ker\Sigma_k.
\]
Taking ranks proves \eqref{S_47}.
Thus, both rank conditions hold for
the same integer $k=s$.
\end{proof}
The following corollary establishes the order-\(r\) observer criterion directly from the design rank conditions.
\begin{corollary}\label{cor:order-r}
A stable functional observer of order $r$ exists if and only if
\begin{equation}\label{eq:order-r-consistency}
\operatorname{rank}
\begin{bmatrix}L&0\\ C&G\\ CA&CF\\ LA&LF\end{bmatrix}
=\operatorname{rank}
\begin{bmatrix}L&0\\ C&G\\ CA&CF\end{bmatrix},
\end{equation}
and, for every $\lambda\in\mathbb C$ with
$\operatorname{Re}\lambda\geq0$,
\begin{equation}\label{eq:order-r-stability}
\operatorname{rank}
\begin{bmatrix}C&G\\ \lambda L-LA&-LF\\ CA&CF\end{bmatrix}
=\operatorname{rank}
\begin{bmatrix}L&0\\ C&G\\ CA&CF\end{bmatrix}.
\end{equation}
These are the order-$r$ conditions of \cite{22}.
\end{corollary}
\begin{proof}
Necessity follows from Lemma~\ref{lem:strongstar-consistency}
with $k=1$ for an observer of order $r$.
For sufficiency, set
$\Sigma_1=\operatorname{col}(Y_1,W_1)$,
$Y_1=[L\ \ 0]$, and
$\Theta_{(L,F,1)}=[LA\ \ LF]$.
Condition \eqref{eq:order-r-consistency} makes
$\Omega_1\Sigma_1=\Theta_{(L,F,1)}$ consistent.
With any generalized inverse $\Sigma_1^{-}$, define
\[
\bar N=\Theta_{(L,F,1)}\Sigma_1^{-}\mathcal J_r,
\qquad
\bar M=(I-\Sigma_1\Sigma_1^{-})\mathcal J_r.
\]
The rank condition \eqref{eq:order-r-stability} is
precisely \eqref{S_47} at $k=1$.
\end{proof}
\begin{remark}
Once \eqref{eqMMD} has been solved and the Hurwitz matrix $N$ has been determined, the coefficients
$\varphi_1,\ldots,\varphi_{k+1}$ are then known, and
\eqref{eq--38} determines the observer matrices $J_i$ uniquely:
\begin{equation}\label{eq--39}
\begin{aligned}
J_1&=\varphi_{k+1},\\
J_i&=\varphi_{k+2-i}
 +\sum_{j=1}^{i-1}N_{k-j+1}J_{i-j},\\[-1mm]
&\hspace{23mm}2\leq i\leq k+1.
\end{aligned}
\end{equation}
For example,
$J_2=\varphi_k+N_kJ_1$ and
$J_3=\varphi_{k-1}+N_kJ_2+N_{k-1}J_1$.
Equivalently, defining
$\mathsf J=\operatorname{col}(J_1,\ldots,J_{k+1})$ and
$\boldsymbol\varphi=\operatorname{col}(\varphi_{k+1},\ldots,\varphi_1)$,
we have
\begin{equation}\label{eq:Phi-gain-recovery}
\Phi(N)\mathsf J=\boldsymbol\varphi,
\qquad \mathsf J=\Phi(N)^{-1}\boldsymbol\varphi,
\end{equation}
where $\Phi(N)$ is the block lower-triangular matrix
whose diagonal blocks are $I_r$ and whose block $(i,j)$,
for $i>j$, is $-N_{k-i+j+1}$ (using one-based block indices).
It is invertible for every choice of $N_i$.
Finally, find $T_i$ from \eqref{eq24c} and set
$J=\operatorname{col}(J_2,\ldots,J_{k+1})$.
This completes the algebraic determination of the observer parameters;
\end{remark}
\subsection{Observer Design}
\label{subsec:observer-construction}
The preceding results establish the algebraic
conditions associated with the proposed observer
structure. We now develop a procedure for
constructing a stable functional observer and
determining its parameters from the observer
equations.
\subsubsection{Asymptotic observer design}
For any $k$ satisfying \eqref{S_46}, the general solution of \eqref{eqMMD} has the form
\cite{RaoMitra1971}:
\begin{align}
\Omega_k
&=\begin{pmatrix}
N_1 & \cdots & N_k &
\varphi_1 & \cdots & \varphi_{k+1}
\end{pmatrix}\nonumber\\
&=\Theta_{(L,F,k)}\Sigma_k^{-}
-Z\left(I-\Sigma_k\Sigma_k^{-}\right),
\label{eq:general-solution}
\end{align}
where $\Sigma_k^{-}$ is any generalized inverse satisfying
$\Sigma_k\Sigma_k^{-}\Sigma_k=\Sigma_k$, and $Z$ is an
arbitrary matrix of compatible dimensions. 
To extract the first $q=kr$ columns, define
$\mathcal J_q=\operatorname{col}(I_q,0_{(k+1)p\times q})$. Then we obtain
$[N_1\;\cdots\;N_k]=\bar N-Z\bar M,$
where
$\bar N=\Theta_{(L,F,k)}\Sigma_k^{-}\mathcal J_q$  and $\bar M=(I-\Sigma_k\Sigma_k^{-})\mathcal J_q$. 
Partition $\bar N=[\bar N_1\;\cdots\;\bar N_k]$ and
$\bar M=[\bar M_1\;\cdots\;\bar M_k]$ into $r$-column blocks.
Then $N=\mathcal A-\mathcal BZ\mathcal C$,
where
\[
\mathcal A=
\begin{pmatrix}
0&I_r&0&\cdots&0\\
0&0&I_r&\cdots&0\\
\vdots&\vdots&\vdots&\ddots&\vdots\\
0&0&0&\cdots&I_r\\
\bar N_1&\bar N_2&\cdots&\bar N_{k-1}&\bar N_k
\end{pmatrix},
\]
with $\mathcal B=\operatorname{col}(0_{(k-1)r\times r},I_r)$
and $\mathcal C=\bar M=[\bar M_1\;\cdots\;\bar M_k]$.
The next lemma connects S$^*$FD
with detectability of the pair $(\mathcal A, \mathcal C)$
\begin{lemma}\label{prop:observer-implies-design-ranks}
If system~\eqref{sys-kratz} is strong$^*$ functionally detectable,
then the  pair $(\mathcal A,\mathcal C)$ is detectable.
\end{lemma}
\begin{proof}
By Lemma~\ref{lem:strongstar-consistency},
conditions \eqref{S_46} and \eqref{S_47}
hold for the same integer $k$.
On the other hand the detectability of the pair  $(\mathcal A,\mathcal C)$ is equivalent to 
\[
\rank\begin{bmatrix}
\lambda I_{kr}-\mathcal A\\
\mathcal C
\end{bmatrix}=kr
\iff
\rank\begin{bmatrix}
\lambda^kI_r-\bar N E_k(\lambda)\\
\bar M E_k(\lambda)
\end{bmatrix}=r,
\]
$\quad \operatorname{Re}\lambda\geq0,$  and where
$
E_k(\lambda)=
\operatorname{col}
(I_r,\lambda I_r,\ldots,\lambda^{k-1}I_r).
$

Assume that the second rank condition
fails for some $\operatorname{Re}\lambda\geq0$.
Then there exists $v\ne0$ such that
\[
\bar M E_k(\lambda)v=0,\qquad
\bar N E_k(\lambda)v=\lambda^kv.
\]
Since
$\bar M=(I-\Sigma_k\Sigma_k^{-})\mathcal J_q$,
define
\[
\omega_{k+1}
=\Sigma_k^{-}\mathcal J_qE_k(\lambda)v.
\]
The first equality yields
\[
\Sigma_k\omega_{k+1}
=\mathcal J_qE_k(\lambda)v.
\]
Moreover, by the definition
$\bar N=\Theta_{(L,F,k)}\Sigma_k^{-}\mathcal J_q$,
\[
\Theta_{(L,F,k)}\omega_{k+1}
=\bar N E_k(\lambda)v=\lambda^kv.
\]

Since
$\Sigma_k=\operatorname{col}(Y_k,W_k)$,
we obtain
\[
W_k\omega_{k+1}=0,\qquad
Y_k\omega_{k+1}=E_k(\lambda)v\ne0.
\]
Together with
$\Theta_{(L,F,k)}\omega_{k+1}=\lambda^kv$,
this gives
\[
\mathcal R_{(L,F,\lambda,k)}
\omega_{k+1}=0.
\]
On the other hand, \eqref{S_46} and
\eqref{S_47} imply
\[
\ker\begin{bmatrix}
\mathcal R_{(L,F,\lambda,k)}\\W_k
\end{bmatrix}
=\ker\Sigma_k.
\]
Hence $\Sigma_k\omega_{k+1}=0$,
contradicting
$Y_k\omega_{k+1}=E_k(\lambda)v\ne0$.
Therefore,
$
\rank\begin{bmatrix}
\lambda I_{kr}-\mathcal A\\
\mathcal C
\end{bmatrix}=kr,
\qquad \operatorname{Re}\lambda\geq0.
$
By the PBH criterion,
$(\mathcal A,\mathcal C)$ is detectable.
\end{proof}
We next state the Lyapunov characterization of
stability for a candidate static output-feedback
matrix $Z$. The stabilizability of
$(\mathcal A,\mathcal B)$ and detectability of
$(\mathcal A,\mathcal C)$ are necessary for the
existence of a stabilizing $Z$, but these
conditions alone are not sufficient in general.
\begin{thm}
\label{thm:SOF-stability}
For a fixed integer $k$ satisfying \eqref{S_46},
a real matrix $Z$ makes
$N=\mathcal A-\mathcal BZ\mathcal C$ Hurwitz
if and only if there exists a symmetric positive
definite matrix $X_L$ such that
\begin{equation}\label{eq:SOF-Lyap}
(\mathcal A-\mathcal BZ\mathcal C)^TX_L
+X_L(\mathcal A-\mathcal BZ\mathcal C)<0.
\end{equation}
\end{thm}
\begin{proof}
For each fixed real $Z$, the matrix
$N=\mathcal A-\mathcal BZ\mathcal C$ is
Hurwitz if and only if there exists
$X_L=X_L^T>0$ satisfying
$N^TX_L+X_LN<0$, by the continuous-time
Lyapunov theorem. Substitution of the
expression for $N$ gives \eqref{eq:SOF-Lyap}.
\end{proof}
The existence of a stabilizing $Z$ is not
asserted by the preceding necessary rank
conditions; it requires feasibility of
\eqref{eq:SOF-Lyap} for some $Z$ and $X_L>0$.
Condition~\eqref{eq:SOF-Lyap} is a bilinear matrix
inequality (BMI) in $X_L$ and $Z$. A stabilizing
solution may be sought using BMI optimization
techniques or iterative linear matrix inequality
(LMI) methods; see, for example,
\cite{36,37}.
The following iterative procedure summarizes the
design of an asymptotic functional observer using
static output feedback.

\begin{enumerate}
\item Verify S$^*$FD. If it does not hold,
stop, since no stable functional observer exists.

\item Set $k=1$ and construct $\Sigma_k$ and
$\Theta(L,F,k)$.

\item Check the consistency condition
\eqref{S_46}. If it fails, increase
$k$ by one and return to Step~2.

\item Compute the generalized-inverse
parametrization \eqref{eq:general-solution}
and construct the matrices
$\mathcal A$, $\mathcal B$, and $\mathcal C$.

\item Seek matrices $Z$ and $X_L=X_L^T>0$
satisfying \eqref{eq:SOF-Lyap}.
If a feasible solution is found, set
\[
N=\mathcal A-\mathcal BZ\mathcal C
\]
and proceed to Step~6. Otherwise, increase
$k$ by one and return to Step~2.

\item Compute the matrices
$N_i$, $\varphi_i$, $J_i$, and $T_i$.
Verify equations
\eqref{eq23} and the stability of $N$.
\end{enumerate}

\subsubsection{Design with prescribed poles}
Fix $k$ and choose real matrices
$N_1^\star,\ldots,N_k^\star\in\mathbb R^{r\times r}$.
They define the companion matrix $N^\star$ in
\eqref{eq:NPcanonical}. Its poles are the roots of
\begin{equation}\label{eq:fixed-poles}
\det\!\left(s^kI_r-\sum_{i=1}^kN_i^\star s^{i-1}\right)=0.
\end{equation}
Choose the coefficients so that $N^\star$ is Hurwitz
and has the desired spectrum. Partition $\Sigma_k=\operatorname{col}(Y_k,W_k)$ as in
\eqref{eq:design-YW}, and put
$\mathcal N^\star=[N_1^\star\ \cdots\ N_k^\star]$.
The remaining design equation is
\begin{equation}\label{eq:fixed-pole-equation}
\Phi_k W_k=\Theta_{(L,F,k)}-\mathcal N^\star Y_k,
\quad \Phi_k=[\varphi_1\ \cdots\ \varphi_{k+1}].
\end{equation}

The following theorem characterizes the solvability
of the observer  equations for prescribed
observer poles.

\begin{thm}
\label{prop:fixed-pole-solvability}
Let the matrices $N_i^\star$ be prescribed such that
the corresponding companion matrix $N^\star$ is Hurwitz.
Then \eqref{eq:fixed-pole-equation} admits a
real solution if and only if
\begin{equation}\label{eq:fixed-pole-rank}
\rank
\begin{bmatrix}
W_k\\
\Theta_{(L,F,k)}-\mathcal N^\star Y_k
\end{bmatrix}
=\rank W_k.
\end{equation}
When this condition holds, all solutions are
parameterized by
\begin{equation}\label{eq:fixed-pole-solutions}
\begin{aligned}
\Phi_k={}&
(\Theta_{(L,F,k)}-\mathcal N^\star Y_k)W_k^{-}\\
&+V(I-W_kW_k^{-}),
\end{aligned}
\end{equation}
where $V$ is an arbitrary matrix of compatible
dimensions, which may be chosen as zero.
The rest of the observer matrices are then obtained from
\eqref{eq--39}, and the matrices $T_i$
from \eqref{eq24c}.
\end{thm}

\begin{proof}
For fixed $\mathcal N^\star$, equation~\eqref{eqMMD}
reduce to
\[
\Phi_kW_k
=\Theta_{(L,F,k)}-\mathcal N^\star Y_k.
\]
This matrix equation is solvable if and only if
condition
 \eqref{eq:fixed-pole-rank} is satisfied.
Under this condition, the general solution
is given by \eqref{eq:fixed-pole-solutions}. The remaining observer
matrices are determined from 
equations~\eqref{eq--39} and \eqref{eq24c}.
\end{proof}

The following iterative procedure summarizes
the design of a functional observer with
prescribed stable poles.

\begin{enumerate}
\item Verify S$^*$FD. If it does not hold,
stop, since no stable functional observer
exists.

\item Set $k=1$ and construct $\Sigma_k$,
$\Theta_{(L,F,k)}$, $W_k$, and $Y_k$.
Select the matrices $N_i^\star$ such that
$\mathcal N^\star$ is Hurwitz.

\item Check the rank condition
\eqref{eq:fixed-pole-rank}.
If it fails, increase $k$ by one, select
the prescribed stable poles for the new
observer order, and return to Step~2.

\item Compute $\Phi_k$ from
\eqref{eq:fixed-pole-solutions} and determine
the  matrices $\varphi_i$, $J_i$,
and $T_i$ using \eqref{eq--39} and
\eqref{eq24c}.
\end{enumerate}
\section{Numerical Example}
\label{sec:published-drive}
As a representative physical example, we consider a
two-mass drive consisting of a DC motor coupled to an
inertial load through a flexible shaft \cite{AstromMurrayMotorDrive}-\cite{ExperimentalTwoMass2023}
 The mechanical
parameters are taken from the experimentally identified
values reported in~\cite{ExperimentalTwoMass2023}.
The model is extended to account
for the first-order dynamics of the voltage actuator
and to include the absolute angular position of
the load as an additional state.
Let $\theta_m(t)$ and $\theta_L(t)$ denote the angular
positions (rad) of the motor and load shafts, respectively,
and let $\omega_m(t)=\dot\theta_m(t)$ and
$\omega_L(t)=\dot\theta_L(t)$ be their angular velocities
(rad/s). The shaft torsional deformation is
$\delta(t)=\theta_m(t)-\theta_L(t)$, with
$\dot\delta(t)=\omega_m(t)-\omega_L(t)$.

The state vector is defined by
\[
x(t)=\operatorname{col}\bigl(
i_a(t),\omega_m(t),\delta(t),
\omega_L(t),v_a(t),\theta_L(t)\bigr),
\]
where $i_a(t)$ is the armature current (A) and
$v_a(t)$ is the armature voltage (V).
The known input $u(t)$ is the voltage command (V),
applied through a first-order actuator with gain $K_a$
and time constant $\tau_a$. The unknown input $d(t)$
represents the external load torque (N\,m).
The electromagnetic torque, back electromotive force,
and elastic shaft torque are given by
$K_ti_a(t)$, $K_e\omega_m(t)$, and $k_s\delta(t)$,
respectively.

The resulting system is described by
\begin{equation}\label{eq:exp-drive-model}
\begin{aligned}
L_a\dot i_a(t)
&=-R_ai_a(t)-K_e\omega_m(t)+v_a(t),\\
J_m\dot\omega_m(t)
&=K_ti_a(t)-b_m\omega_m(t)-k_s\delta(t),\\
\dot\delta(t)
&=\omega_m(t)-\omega_L(t),\\
J_L\dot\omega_L(t)
&=k_s\delta(t)-b_L\omega_L(t)-d(t),\\
\tau_a\dot v_a(t)
&=-v_a(t)+K_au(t),\\
\dot\theta_L(t)
&=\omega_L(t).
\end{aligned}
\end{equation}
Here $R_a$ ($\Omega$) and $L_a$ (H) are the armature
resistance and inductance; $K_e$ (V\,s/rad) and
$K_t$ (N\,m/A) are the back-EMF and torque constants;
$J_m$ and $J_L$ (kg\,m$^2$) are the motor and load
inertias; $b_m$ and $b_L$ (N\,m\,s/rad) are the
viscous friction coefficients; and $k_s$ (N\,m/rad)
is the shaft torsional stiffness. The actuator
parameters are its time constant $\tau_a$ (s)
and static gain $K_a$ (V/V).
The subscripts $m$ and $L$ refer to the motor and
load, respectively.
The experimentally identified mechanical parameters
reported in~\cite{ExperimentalTwoMass2023} are
\begin{equation}\label{eq:exp-drive-parameters}
\begin{gathered}
J_m=1.5602\times10^{-3},\quad J_L=1.3966\times10^{-3},\\
k_s=19.4840,\quad b_m=3.7167\times10^{-3},\\
b_L=3.5530\times10^{-3}.
\end{gathered}
\end{equation}
For the augmented electrical and actuator dynamics,
we take $R_a=0.1\,\Omega$, $L_a=0.01$ H,
$K_e=0.1$ V\,s/rad, $K_t=0.1$ N\,m/A,
$\tau_a=0.25$ s, and $K_a=1$.
In particular, $K_e$ and $K_t$ are chosen to have
the same numerical value in SI units, consistently
with the electromechanical coupling of a DC motor.
 The matrices of this system are:
\begin{equation}\label{eq:exp-drive-matrices}
\begin{gathered}
A=\begin{pmatrix}
-10&-10&0&0&100&0\\
64.0943&-2.3822&-12488.14&0&0&0\\
0&1&0&-1&0&0\\
0&0&13950.74&-2.5440&0&0\\
0&0&0&0&-4&0\\
0&0&0&1&0&0
\end{pmatrix},\\
B=\operatorname{col}(0,0,0,0,K_a/\tau_a,0),\qquad\\
F=\operatorname{col}(0,0,0,-1/J_L,0,0),\\
C=\begin{pmatrix}0&1&0&0&0&0\\0&0&0&1&0&0\\0&0&1&0&0&0\end{pmatrix},\\
G=0_{3\times1},\quad \text{and} \quad L=\begin{pmatrix}1&0&0&0&0&0\end{pmatrix}.
\end{gathered}
\end{equation}
 The measured outputs are $y=\operatorname{col}(\omega_m,\omega_L,\delta)$ and the variable to be estimated is the current $z=i_a$..
This system is neither strongly observable nor
strongly detectable, and then is not strongly*
detectable. However,
the functional $z=i_a$ is  (S$^*$FD). On the other hand  the two
conditions of Corollary~\ref{cor:order-r}
are not satisfied for $r=1$. Then we augment $k$ to $k=2$ corresponding to an observer of dimension $2$. Then we 
apply the pole-placement procedure. Following the companion structure
\eqref{eq:NPcanonical}, we prescribe
\[
N^\star=
\begin{pmatrix}
0&1\\
N_1^\star&N_2^\star
\end{pmatrix}.
\]
Assigning the poles at $-3$ and $-5$
gives $N_1^\star=-15$ and $N_2^\star=-8$.
The functional observer has the following form:
\begin{equation}\label{eq:exp-drive-observer}
\dot{\xi}(t)=N^\star \xi(t)+Jy(t)+B_0u(t),\\
\widehat z(t)=P\xi(t)+J_1y(t),
\end{equation}
with
$N^\star=\begin{pmatrix}0&1\\-15&-8\end{pmatrix},$ $P=\begin{pmatrix}1&0\end{pmatrix},$\\
$J=\begin{pmatrix}
-9.864156&0&-1169.04\\
39.388253&0&4481.32
\end{pmatrix},$
$B_0=\begin{pmatrix}0\\400\end{pmatrix},$\\
and $J_1=\begin{pmatrix}-0.093612&0&0\end{pmatrix}.$
The obtained matrix $T$  is given by 
\[
T=\begin{pmatrix}
1&0.093612&0&0&0&0\\
-4&-0.358846&0&0&100&0
\end{pmatrix}.
\]
For the simulation, the applied control is $u(t)=0.5\,H(t)\,\mathrm{V}$,
where $H(t)$ denotes the Heaviside step function., and the unknown load torque
$d(t)=0.03\sin(2t)+0.02\cos(7t)$ N\,m.
 The initial conditions are
$x(0)=(0.3,0.2,0.01,-0.15,0.5,1)^\top$ and
$\xi(0)=(0,0)^\top$. 
Figure~\ref{fig:exp-drive-inputs} shows the applied inputs.
Figures~\ref{fig:exp-drive-current} and~\ref{fig:exp-drive-error}
show the simulated current and its signed estimation error.
 The  absolute-angle
mode persists, while the functional estimation error
converges to zero.
\begin{figure}[t]
\centering\includegraphics[width=\columnwidth]{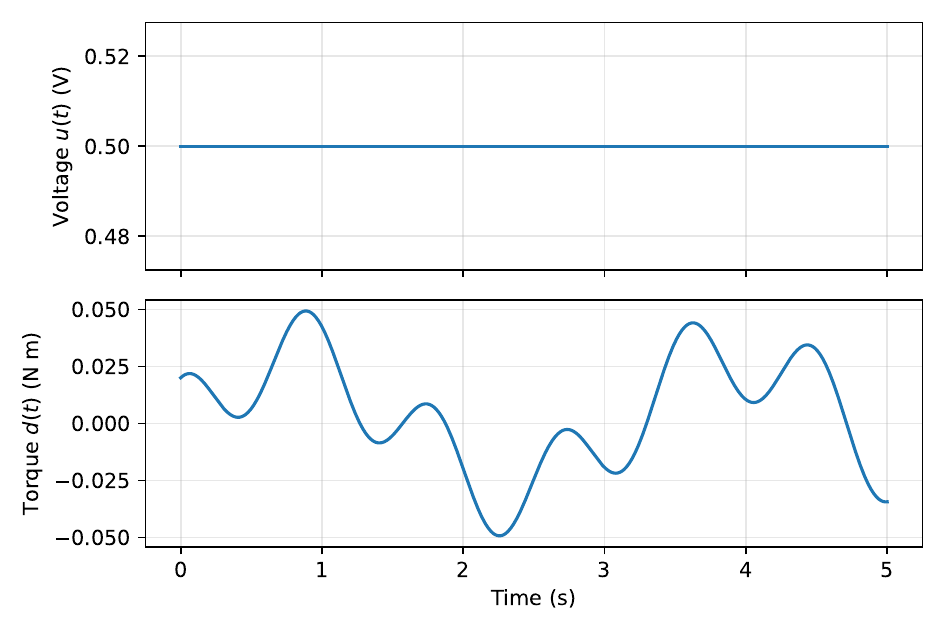}
\caption{Applied DC voltage step $u(t)=0.5H(t)$ and time-varying unknown load torque $d(t)$ (separate vertical scales).}\label{fig:exp-drive-inputs}
\end{figure}
\begin{figure}[t]
\centering\includegraphics[width=\columnwidth]{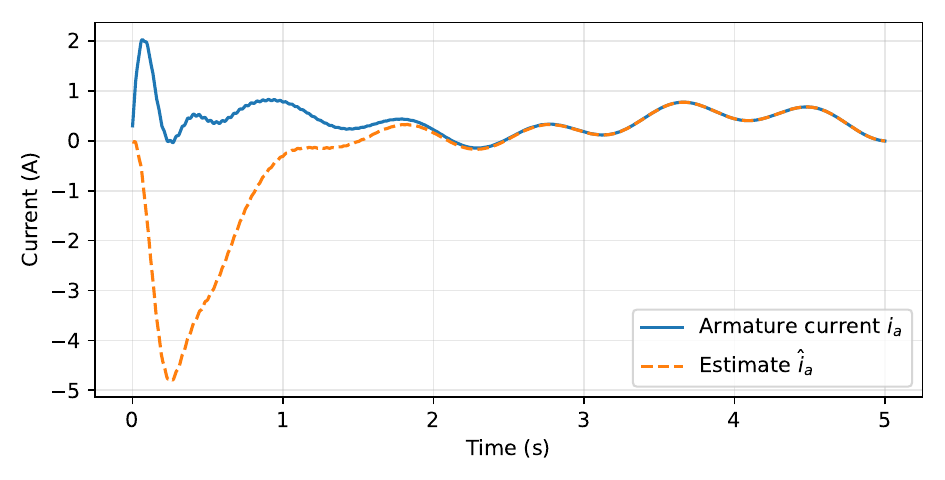}
\caption{Armature current and its functional estimate for the augmented experimental two-mass drive.}\label{fig:exp-drive-current}
\end{figure}
\begin{figure}[t]
\centering\includegraphics[width=\columnwidth]{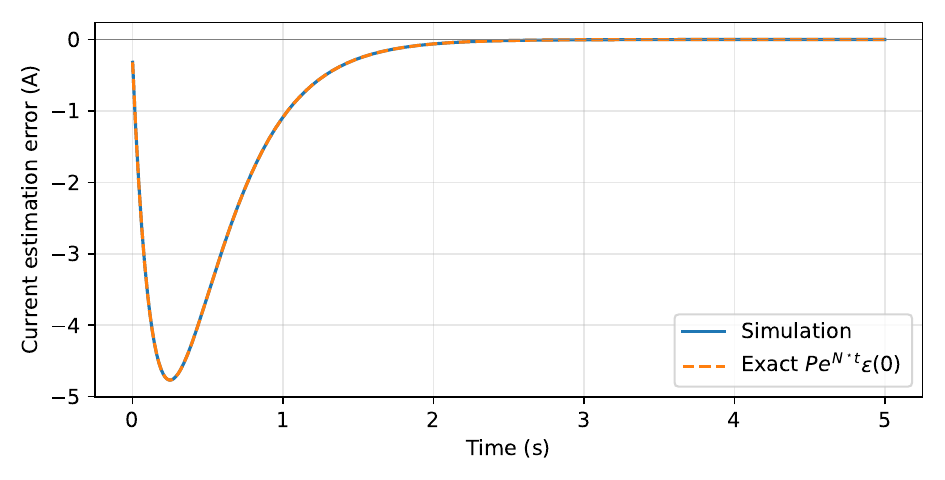}
\caption{Signed current estimation error for the companion-form observer: numerical simulation and exact error dynamics with poles $-3$ and $-5$.}\label{fig:exp-drive-error}
\end{figure}

\section{Conclusion}
This paper has investigated SFO, SFD, and
S$^*$FD for continuous-time linear systems
with unknown inputs affecting both the state
and measurement equations. Extending Kratz's
finite algebraic construction, necessary and
sufficient algebraic conditions have been
established for characterizing these properties
without differentiating the unknown inputs.
The relationship between S$^*$FD and the
existence of stable functional observers has
also been examined. Furthermore, a constructive
observer-design procedure has been developed.
Particular cases and a numerical process example
have been presented to illustrate the theoretical
results and the effectiveness of the proposed
design procedure. 
Future research will investigate the design
of functional observers of arbitrary order
$q\geq r$, removing the restriction that
the observer order be a multiple of $r$,
with the aim of determining the minimum
achievable order.

\end{document}